\documentstyle[sprocl,epsf]{article}

\def\Journal#1#2#3#4{{#1} {\bf #2}, #3 (#4)}

\def\PRD{{\em Phys. Rev.} D}

\def\be{\begin{equation}}
\def\ee{\end{equation}}
\def\bea{\begin{eqnarray}}
\def\eea{\end{eqnarray}}

\begin{document}

\begin{flushright}
 PITHA 00/25\\
 TPR-00-19\\
 hep-ph/0010208\\
 13 October 2000\\
\end{flushright}
\vskip 1cm

\title{RENORMALONS AND POWER CORRECTIONS\footnote{To be published 
in the Boris Ioffe Festschrift ``At the Frontier of Particle
Physics/Handbook of QCD'', edited by M. Shifman (World Scientific, 
Singapore, 2001)}}

\author{M.~BENEKE}

\address{Institut f\"ur Theoretische Physik E, RWTH Aachen,\\ 
Sommerfeldstr.~28, D-52074 Aachen, Germany}

\author{V.M.~BRAUN}

\address{Institut f\"ur Theoretische Physik, Universit\"at
   Regensburg, \\ Universit\"atsstr.~31, D-93053 Regensburg, Germany
}

\maketitle\abstracts{Even for short-distance dominated observables 
the QCD perturbation expansion is never complete. The divergence 
of the expansion through infrared renormalons provides formal evidence
of this fact. In this article we review how this apparent failure 
can be turned into a useful tool to investigate power corrections 
to hard processes in QCD.}


\tableofcontents

\newpage


\section{Introduction}
\label{sec:introduction}

Short-distance phenomena in strong-interaction physics are usually 
described in the framework of a perturbation expansion. The need for 
high precision is one reason why we may want to reach beyond the 
limitations of this framework. But there are also interesting
questions regarding the structure and meaning of the perturbation 
series itself which naturally lead us to investigate long-distance, 
non-perturbative aspects of hard processes in QCD. The purpose of this
article is to exhibit this relation of perturbative and
non-perturbative physics, and the kind of phenomenology it has 
entailed. Some of the details which we have to skip here can 
be found elsewhere.\,\cite{physrep}

To state this more clearly we consider an example from deep-inelastic 
neutrino-nucleon scattering. In the parton model the 
Gross-Llewellyn-Smith (GLS) sum rule for the structure function $F_3$ 
expresses the fact that the proton consists of three valence quarks. 
In QCD there are finite corrections to this sum rule, so that 
\begin{eqnarray}
\label{eq:gls}
\lefteqn{I_{\rm GLS}\equiv 
\frac{1}{2}\int_0^1 \!dx \,\,(F_3^{\nu p}+F_3^{\bar{\nu}p})(x,Q) 
=}
\nonumber\\
&=&
3\left[1
-\frac{\alpha_s}{\pi}
-3.58 \left(\frac{\alpha_s}{\pi}\right)^2
-18.98 \left(\frac{\alpha_s}{\pi}\right)^3 
-\ldots -\frac{C_2}{Q^2} +O(1/Q^4)\right],
\end{eqnarray}
where $\alpha_s=\alpha_s(Q)$ and the leading higher-twist correction 
$C_2$ has been estimated 
to be $0.1\,\mbox{GeV}^2$. Conventional parlance would say that the 
sum rule has a perturbative contribution (the series in $\alpha_s/\pi$) and 
a non-perturbative one, but this is an imprecise characterization 
of Eq.~(\ref{eq:gls}), because the perturbative series diverges 
for any value of $\alpha_s$. What is the numerical value of the 
``perturbative contribution''?

There are several reasons for why the series expansion might
diverge. The divergence that has most implications is known as the 
infrared (IR) renormalon.\,\cite{GN74,L77,tH77} (The other known sources 
of divergence are: instan\-ton-anti\-instanton pairs, which are suppressed 
in QCD; ultraviolet renormalons, which -- at least in principle --  
can be disposed of for practical purposes. We briefly discuss 
ultraviolet renormalons in Sect.~\ref{sec:uvrenormalon}.) 
For the GLS sum rule it can be shown that 
the coefficients $r_n$ of $\alpha_s^{n+1}$ diverge (due to IR
renormalons) for large $n$ as 
\begin{equation}
\label{eq:div}
r_n \propto \left(\frac{2 \beta_0}{p}\right)^n n!\, n^b,
\end{equation}
where $p=2$, $\beta_0=(11-2N_f/3)/(4\pi)$, $N_f$ the number of
massless quarks, and $b$ some constant which we assume here to be zero for
simplicity. A divergent series expansion 
is a useful approximation if it is asymptotic to the quantity which 
it represents. QCD perturbative expansions have never been proven to be 
asymptotic, but it is a good idea to proceed with this assumption 
on good faith. Assuming furthermore that 
\begin{equation}
\left|I_{\rm GLS}-3 \sum_{n=-1}^N r_n \alpha_s^{n+1}\right| < K_{N+1} 
\alpha_s^{N+2}
\end{equation}
with $K_n \propto r_n$ the best approximation occurs at $N=N_0 \sim 
|p|/(2 \beta_0\alpha_s)$ and 
\begin{equation}
\label{eq:exp}
\left[K_{N_0}\alpha_s^{N_0+1}\right]_{\rm min}
 \sim e^{-|p|/(2 \beta_0\alpha_s)} \sim 
\left(\frac{\Lambda}{Q}\right)^p.
\end{equation}
With $p=2$ this is of the same order of magnitude as the power
correction $C_2/Q^2$ in Eq.~(\ref{eq:gls}). Such power corrections 
are referred to as ``non-perturbative'', because they are
exponentially small in the strong coupling $\alpha_s(Q)$. Without a 
summation prescription, however, the numerical value of 
the ``perturbative contribution'' is not unique, although it is
determined to an accuracy $(\Lambda/Q)^2$, which is small, when 
$Q$ is large. It therefore seems that to include these power
corrections consistently we only need to figure out the correct 
prescription to sum the perturbative series and add to the sum 
the higher-twist correction $C_2/Q^2$ -- but this is wrong! Defining 
what $C_2$ is requires great care, and once the power correction 
is properly defined, the summation prescription for the perturbative 
series is {\em implied} by this definition.\,\cite{D84,NSVZ85,Mue85} 
Two important consequences
follow: the perturbative series and power (``non-perturbative'') 
corrections are not independently defined, they are related; if 
the series diverges as in Eq.~(\ref{eq:div}), then there must be a
power correction of order $(\Lambda/Q)^p$, whose precise definition
fixes the ambiguity in defining the perturbative expansion. We can 
use this to obtain some insight into power corrections using nothing 
but the rules of perturbative QCD.

This becomes much clearer, if we take into account the physics origin 
of IR renormalons. The sum of Feynman amplitudes at a given order 
in perturbation theory, which give the 
perturbative expansion of the GLS sum rule, is IR finite and 
depends only on the large scale $Q$. On dimensional grounds the average loop 
momenta must therefore scale with $Q$. However, the coefficient of 
proportionality may depend strongly on the order of perturbation 
theory. Suppose in an $n+1$ loop contribution to $I_{\rm GLS}$ we 
have integrated over all loop momenta but one momentum $k$ and 
that the result of the $n$ loop integrations is proportional 
to $\beta_0^n \ln^n (Q^2/k^2)$. The coefficient $\beta_0^n$ 
appears unmotivated at this stage and will be explained in 
Sect.~\ref{sec:basics}, but for now we may only note that the
scale dependence of the coupling is an obvious source of logarithms.  
Then the dominant contributions to the final integral over $k$ 
come from $k\gg Q$ and $k\ll Q$, because of the large logarithmic 
enhancements in these regions. (The contribution from 
large $k$ is related to ultraviolet renormalons and we ignore it in 
the following.) If $I(k)$, the integrand for the final loop
integration, goes as $I(k) \sim k^{p-4}$ for $k\ll Q$, we find 
\begin{equation}
\label{eq:smallk}
r_n \sim \int d^4 k \,I(k) \,\beta_0^n \ln^n (Q^2/k^2) 
\sim \left(\frac{2 \beta_0}{p}\right)^n n!
\end{equation}
as in Eq.~(\ref{eq:div}) with typical $k\sim Q e^{-n/p}$. The
contribution to $r_n\alpha_s^{n+1}$ from $k<\mbox{few}\cdot \Lambda$ 
is again of order $(\Lambda/Q)^p$. The rules of perturbative 
QCD cannot be assumed to account for this small loop momentum region. 
This is not a problem in conventional 1-loop or 2-loop 
calculations, since the power-suppressed contribution from small 
momenta is much smaller than the dominant contribution of order 
$\alpha_s$ or $\alpha_s^2$. But since we are now interested in 
such small power-suppressed effects, we should better consider the 
small loop momentum contributions as part of the low-energy 
matrix elements of higher-twist operators in the operator product 
expansion (OPE) of the GLS sum rule.

We therefore define $C_2$ as the matrix element of the relevant 
twist-4 opera\-tor (strictly speaking, as the product of 
the matrix element and a coefficient function) which includes all low 
momentum contributions with $k < \mu$ for some $\mu\ll Q$, but larger 
than $\Lambda$. The contribution from $k<\mu$ has to be accordingly 
subtracted from the perturbative expansion. Since the higher-twist 
operator is quadratically ultraviolet divergent, we know that 
$C_2 = c\mu^2+O(\Lambda^2)$ for $\mu\gg\Lambda$ and hence we 
can rewrite Eq.~(\ref{eq:gls}) as 
\begin{equation}
\label{eq:sub}
I_{\rm GLS}=
3\bigg[\bigg\{1-\sum_{n=0}^\infty r_n\alpha_s^{n+1}+
\frac{c\mu^2}{Q^2}\bigg\} -\frac{C_2}{Q^2} +O(1/Q^4)\bigg]
\end{equation}
with the subtraction of low momentum regions taken into account 
in the curly brackets. For $\mu$ sufficiently large 
compared to $\Lambda$, $c$ itself has a perturbative 
expansion in $\alpha_s$. This expansion is exactly such that 
in large order it cancels the IR renormalon divergence of the 
coefficients $r_n$ so that the two expansions in curly brackets 
combined are convergent -- more precisely, the 
remaining IR renormalon divergence causes a summation ambiguity 
of higher order in the $1/Q$ expansion. By defining $C_2$ 
accurately, we succeeded in summing the divergent series 
by eliminating the divergence altogether! However, the 
short-distance ($k>\mu$, ``perturbative'') and long-distance 
($k<\mu$, ``non-perturbative'') terms in Eq.~(\ref{eq:sub}) 
depend separately on $\mu$ and remain related just as before. 
But we now see that it is really the small-$k$ behavior of the  
1-loop integrand $I(k)$ rather than factorial divergence 
which determines the magnitude of the power correction.  

The analogy with conventional scale dependence may help 
understanding why the IR cutoff dependence of Feynman amplitudes can give
information on non-perturbative power corrections. The former scale 
dependence relates different orders in perturbation theory and 
is often used to estimate the size of unknown higher-order 
corrections; the $\mu$-dependence in Eq.~(\ref{eq:sub}) relates different 
orders in the $1/Q$ expansion and may be used to estimate the 
size of the first power corrections. These estimates are
parametrically correct, but quantitatively only indicative. The 
important point is that if the scale dependence is large, then 
so must be the higher-order correction in $\alpha_s$, or $1/Q$. (It is
often stated that 
the perturbative expansions are scale-independent to all orders of 
perturbation theory. Because the expansion is divergent this 
is only formally true. Any attempt to interpret the series 
numerically introduces the kind of scale dependence or prescription 
dependence exhibited in Eq.~(\ref{eq:sub}).)

There exist several ways of making use of this connection between 
IR renormalons and power corrections:
\begin{itemize}
\item[-] {\em formal}: if power corrections can be analyzed with 
operator product expansion methods, the same methods can be used 
to determine the IR renormalon divergence.\,\cite{P78,P79,Mue85} 
This goes as far as 
fixing all parameters of Eq.~(\ref{eq:div}), including subleading 
corrections in $1/n$, but excepting the overall constant of 
proportionality, since most of the structure of Eq.~(\ref{eq:div}) 
is determined by logarithms which can be controlled by OPE and 
renormalization group methods. This is perhaps of less interest 
phenomenologically, except to remind us that combining perturbative 
expansions with higher-order terms in the OPE is subtle unless 
we can argue that the matrix elements of higher-dimension operators 
are much larger than the low momentum contributions in perturbative 
Feynman amplitudes.
\item[-] {\em qualitative/scaling}: here we begin with the divergence 
of the perturbative expansion and deduce from it the scaling with $Q$  
of power corrections. Some power corrections can be missed in 
this way, but usually there is an identifiable reason for this.  
The great advantage of this method is that the quantity does not have 
to admit an operator product expansion, the only requirement being that 
it has a short-distance scale, i.e. a perturbative expansion to begin with. 
In general this is a poor substitute 
for a full understanding of power corrections in terms of operators, 
but in some cases this is the only method known to this date. Rather 
than speaking of divergent perturbative expansions, we could directly 
investigate the small momentum behavior of Feynman amplitudes. This 
makes apparent the close relation of this approach with the methods 
of perturbative infrared factorization, but now extending this notion 
beyond the study of logarithmic collinear and soft infrared 
sensitivity at the leading power in the hard scale. The Feynman 
amplitudes are implicitly presumed to indicate the correct scaling 
behavior of non-perturbative corrections.  
\item[-] {\em quantitative}: this is the most interesting, but also most 
delicate of all applications. From what has been said it seems 
impossible to obtain quantitative information on power corrections 
by perturbative methods. We cannot compute the small momentum contribution 
of arbitrarily complicated Feynman amplitudes, and
even if we could, this would be of no use since this does not give the 
correct non-perturbative result. However, we can imagine a situation 
in which the subtraction term in curly brackets in Eq.~(\ref{eq:sub}) 
cancels almost completely all higher-order terms in the original 
pertubative expansion (say, for $n>1$) for some value of $\mu$, because 
the higher-order terms are already dominated by small loop momentum. 
If the value of $\mu$ at which this occurs (say, $\mu=1\,$GeV) is 
numerically large compared to $\Lambda$, the power correction $C_2/Q^2$ 
in Eq.~(\ref{eq:sub}) might be numerically dominated by the first term 
in the expansion of $c \mu^2$ in $\alpha_s(\mu)$. In this case 
Eq.~(\ref{eq:sub}) is well approximated by a perturbative expansion 
truncated at $n=1$ together with a power correction whose numerical 
coefficient is predicted by IR renormalons or perturbative IR 
contributions! This is of course a rather idealized situation but 
we shall see that this logic provides an explanation of the sometimes 
puzzling success of models of power corrections based on perturbative 
infrared sensitivity. Since the power correction is really a 
parametrization of perturbative contributions, though originating at 
scales much smaller than the hard scale $Q$, we should expect that 
the ``non-perturbative'' power correction decreases as more terms are 
added to the perturbative expansion. 
\end{itemize}
The outline of this review is as follows: 

In Sect.~\ref{sec:basics} we show how the IR renormalon divergence 
can be characterized with OPE methods. This is then illustrated by 
computing a set of fermion bubble diagrams, though in general we shall 
try to free the notion of renormalons from its association with this 
rather special set of diagrams. The step to quantities without an 
OPE is made in this section, appealing to the more general concept of 
perturbative infrared sensitivity as already discussed in this 
introduction.

Section~\ref{sec:applications} 
concentrates on phenomenological applications. In our 
opinion ideas based on, or inspired by, IR renormalons have had the most 
important impact on our understanding of power corrections to hadronic 
event shape measures; on perturbative effects in heavy 
quark decays and production  
due to the clarification of the role of heavy quark mass definitions; 
on the modelling of twist-4 corrections in deep-inelastic nucleon 
structure functions. These three applications are discussed in some 
detail. Others can only be briefly summarized.

Section~\ref{sec:uvrenormalon} treats rather 
briefly the physics of ultraviolet renormalons. 
We conclude in Sect.~\ref{sec:conclusion}.

\section{Infrared Renormalons - Basic Concepts}
\label{sec:basics}

\subsection{The Borel Plane}

Renormalon divergence is often discussed in terms of the corresponding
singularities in the Borel plane. We briefly introduce the relevant 
concepts.

Given a quantity $R$ and its series expansion, we define the Borel 
transform $B[R](t)$ of the series by
\begin{equation}
\label{defborelt}
R\sim\sum_{n=0}^\infty r_n\alpha_s^{n+1} \,\,\Longrightarrow \,\,
B[R](t) = \sum_{n=0}^\infty r_n\,\frac{t^n}{n!}.
\end{equation}
If $B[R](t)$ has no singularities for real 
positive $t$ and does not increase too 
rapidly at positive infinity, we can define the 
Borel integral ($\alpha_s$ positive) as 
\begin{equation}
\label{borelint}
\tilde{R} = \int_0^\infty \,dt\,e^{-t/\alpha_s}\,B[R](t),
\end{equation}
which has the same series expansion as $R$. In QCD the Borel integral 
does not exist, since IR renormalons generate singularities on the 
integration contour. The non-existence of the integral is not a 
serious concern, however, since we would not have expected the 
Borel integral to equal $R$ anyway, because of non-perturbative, 
power-suppressed effects. From Eq.~(\ref{borelint}) we see that the 
effect of singularities at finite $t$ on $R$ is also
power-suppressed. Unless $b$ is a negative integer, the 
correspondence between factorial divergence and singularities is 
as follows:
\begin{equation}
\label{div}
r_n=K a^n\Gamma(n+1+b)\,\,\Longleftrightarrow \,\,
B[R](t) = \frac{K\Gamma(1+b)}{(1-a t)^{1+b}}.
\end{equation}
Because of this the divergent behavior of the original series is 
encoded in the singularities of its Borel transform. Hence, divergent 
behavior is often referred to through poles/singularities in 
the Borel plane. This language is particularly convenient for 
subleading divergent behavior. Note that larger $a$, i.e. faster 
divergence, leads to singularities closer to the origin $t=0$ of the 
Borel plane.

\begin{figure}[t]
   \vspace{-3.2cm}
   \epsfysize=21cm
   \epsfxsize=14cm
   \centerline{\epsffile{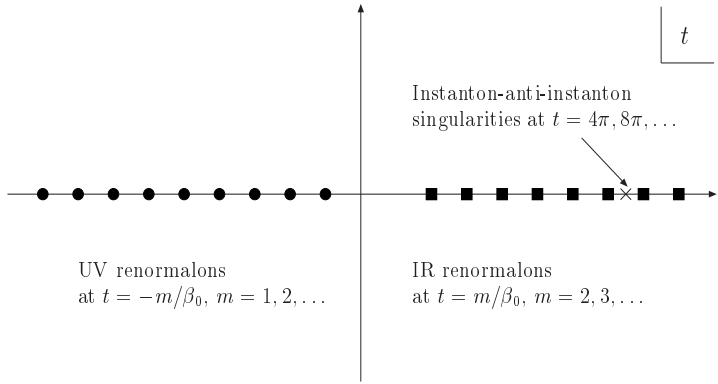}}
   \vspace*{-12.2cm}
\caption[dummy]{\small Singularities in the Borel plane of 
$\Pi(Q^2)$. The singular points are shown, 
but not the cuts attached to each of them. 
\label{fig1}}
\end{figure}

To illustrate these concepts and survey the known singularities, we 
consider the correlation functions of two vector currents $j_\mu=
\bar{q}\gamma_\mu q$ of massless quarks
\begin{equation}
\label{currentcorr}
(-i)\int\,d^4x\,e^{-i q x}\,\langle 0|T\,(j_\mu(x) j_\nu(0))|0\rangle 
= \left(q_\mu q_\nu-q^2 g_{\mu\nu}\right)\,\Pi(Q^2)
\end{equation}
with $Q^2=-q^2$. The singularities in the Borel plane are shown in 
Fig.~\ref{fig1}. How general is this picture? In QCD, ultraviolet renormalon 
and instanton-antiinstanton singularities always occur at the
locations indicated in the figure, independent of the particular 
observable, because the physics they represent (ultraviolet behavior
and vacuum structure, respectively) is universally the same. The
location of IR renormalon singularities depends on the specific 
observable. For observables derived from off-shell correlation
functions, such as $\Pi(Q^2)$ and the GLS sum rule, IR renormalon 
singularities occur at integer multiples of $1/\beta_0$, because 
the OPE is an expansion in even powers of the hard scale. For
observables derived from on-shell correlation functions, one can have 
power corrections suppressed by only one power of the hard scale. 
These would be related to a singularity at $t=1/(2 \beta_0)$. In
general, one can construct infrared finite observables, which 
are arbitrarily infrared-sensitive.\,\cite{MW95} IR renormalons can 
then occur at any $t$ and arbitrarily close to 0. 

\subsection{IR Renormalons and the Operator Product Expansion}

{From} Eq.~(\ref{eq:smallk}) we learned that factorial divergence 
arises if the typical loop momentum $k\to 0$ as the order of 
perturbation theory increases. We take this to be the definition of 
what we mean by ``IR renormalon'', i.e. an IR renormalon is a
singularity of the Borel transform that is eliminated if all loop 
momenta are restricted to $k>\Delta$ no matter how small $\Delta$ 
is. By definition the analysis of IR renormalons is the analysis 
of small momentum contributions to Feynman amplitudes. This appears 
complicated since IR renormalons refer to large orders in perturbation
theory and so it seems that we would need to investigate the 
infinite set of Feynman diagrams. We shall now see that the
factorization properties of Green functions strongly constrain the 
form IR renormalon divergence can take.\,\cite{P79,Mue85}

For the remainder of this subsection we restrict ourselves to 
observables derived from off-shell (Euclidean) correlation 
functions; the function $\Pi(Q)$ defined in Eq.~(\ref{currentcorr}) 
will serve as an example. The factorization of small loop momentum 
regions is done for us by the operator product 
expansion\,\cite{Wil69,CHM72,SVZ79} (OPE), which for 
$\Pi(Q)$ reads
\begin{equation}
\label{eq:ope}
\Pi(Q) = C_0(\alpha_s,Q/\mu) + \frac{1}{Q^d}\,C_d(\alpha_s,Q/\mu)\,
\langle \frac{\alpha_s}{\pi} G^2\rangle(\mu) + O(1/Q^6),
\end{equation}
where $d=4$ and the leading power correction is given by the gluon 
condensate. The conventional way of interpreting this formula is 
to take $C_0$ as the series computed with the standard rules of 
QCD perturbation theory. This includes the small loop momentum regions
which give rise to IR renormalons. It is conceptually more
satisfactory to include these regions into the definition of the 
vacuum condensates -- the OPE guarantees that this can always be 
done. The series expansion of $C_0$ then differs from the standard 
one and is free from IR renormalons. Both $C_0$ and the vacuum
condensates are in this case separately well-defined. 
In the following we shall 
adhere to the conventional interpretation, but we will return to 
the second more appealing one later in this section.

Since any loop momentum region with $k<\Delta$ can be 
absorbed into a series of vacuum expectation values, the IR renormalon
contribution to $C_0$ must take the factorized form 
\begin{equation}
\label{eq:irrope}
C_0^{\rm IR}(\alpha_s,Q/\mu) = \frac{1}{Q^d}\,C_d(\alpha_s,Q/\mu)\,
\mu^d M(\alpha_s) + O(1/Q^6),
\end{equation}
with the same coefficient function $C_d$ as in Eq.~(\ref{eq:ope}) and 
$M(\alpha_s)$ a (dimensionless) 
perturbative series independent of $Q$. The 
coefficient function of the gluon condensate, $C_d$, satisfies the
renormalization group equation
\begin{equation}
\left(\mu^2\frac{\partial}{\partial\mu^2}+\beta(\alpha_s)
\frac{\partial}{\partial\alpha_s} - \frac{\gamma(\alpha_s)}{2}
\right) C_d(\alpha_s,Q/\mu) = 0,
\end{equation}
where $\alpha_s=\alpha_s(\mu)$ and $\gamma(\alpha_s)$ is the
anomalous dimension of the gluon condensate. Since $C_0$ and 
hence $C_0^{\rm IR}$ is independent of $\mu$, this implies that 
\begin{equation}
\label{eq:me}
\left(\beta(\alpha_s)\frac{\partial}{\partial\alpha_s} + 
\frac{d+\gamma(\alpha_s)}{2}\right) M(\alpha_s) = 0.
\end{equation}
Here it is used that $M$ being independent of $Q$ and dimensionless 
cannot depend explicitly on $\mu$. The IR renormalon divergence 
must be contained entirely in $M$. We make the ansatz
\begin{equation}
\label{eq:ansatz}
M(\alpha_s) = \sum_n \alpha_s^{n+1} K (a\beta_0)^n n! \,n^b 
\left(1+\frac{s_1}{n}+O(1/n^2)\right)
\end{equation}
and insert it into Eq.~(\ref{eq:me}). We use $\beta(\alpha_s)=-\beta_0
\alpha_s^2-\beta_1\alpha_s^3-\ldots$ and $\gamma(\alpha_s)=
\gamma_0\alpha_s+\ldots$ and after shifting the summation index $n$
in some terms and expanding in $1/n$, we obtain
\begin{eqnarray}
\label{eq:par1}
0 &=& \sum_n \alpha_s^{n+1} K (a\beta_0)^n n! \,n^b 
\bigg[\frac{d}{2}-\frac{1}{a}
\nonumber\\
&&\hspace*{1cm}+\,\frac{1}{n}\bigg(\frac{b}{a}-\frac{\beta_1}{a^2\beta_0^2}
+\frac{\gamma_0}{2 a\beta_0}+\bigg(\frac{d}{2}-\frac{1}{a}\bigg) s_1\bigg) 
+O(1/n^2)\bigg]
\end{eqnarray}
This implies that either $K=0$, in which case there is no factorial 
divergence, or otherwise we must have
\begin{equation}
\label{eq:ab}
a=\frac{2}{d},\qquad b=\frac{d\beta_1}{2\beta_0^2}-
\frac{\gamma_0}{2\beta_0}.
\end{equation}
In the case of the gluon condensate we have $d=4$ and $\gamma_0=0$, 
in which case we reproduce the result originally derived by 
Mueller\,\cite{Mue85} in a different way. Note that $s_1$ and 
higher-order terms are also determined by the $\beta$-function 
and the anomalous dimension.\,\cite{Ben93b}

It is important to stress the general nature of this result, since 
it applies to all observables for which we know the structure of power
corrections. There may be power corrections and no corresponding 
IR renormalons, i.e. $K=0$, even if the higher-dimension operator 
has a power divergence (and hence a perturbative contribution). This 
can occur in conformal theories such as supersymmetric Yang-Mills 
theory with four unbroken supersymmetries. However, {\em if} the divergence 
occurs, the $n$-dependence is completely determined by renormalization
constants, the perturbative expansion of $C_d$ and the truly 
``non-perturbative'' parameter $K$.\,\cite{Gru93a,Ben93b} There is 
a common misconception that the involvement of the non-abelian 
$\beta$-function coefficient $\beta_0$ in the location of the 
renormalon singularity is only a ``conjecture''. This misconception is
based on the set of fermion bubble diagrams (see the next subsection),
which gives only part of $\beta_0$. It is indeed difficult to identify
diagrammatically the factor $\beta_0^n$, but the renormalization group
allows us to bypass the difficulty. This should not be too surprising 
since the origin of renormalons is large logarithms. Hence we can 
prove the factor $\beta_0^n$, if the renormalon singularity exists, 
but we cannot rigorously prove that renormalons exist in QCD, 
for we cannot exclude that $K=0$ for some mysterious (hence
entirely improbable) reason. 

Suppose we had excluded all loop momenta $k<\Delta$ from the coefficient 
functions. The IR renormalon would then reappear as the coefficient 
of an ultraviolet power divergence of the gluon condensate, although 
the condensate being defined non-perturbatively 
with the cutoff $k<\Delta$, there is no need to separate this
divergence from the remaining condensate. It is however instructive 
to see how this works in an example. We cannot compute condensates 
analytically in QCD, but the non-linear $O(N)$ $\sigma$-model 
in two space-time dimensions  
provides a nice toy model,\,\cite{D82,D84,NSVZ84,NSVZ85,Ter87,BBK98} 
which is solvable in a $1/N$-expansion. 
As QCD it has only massless 
particles in perturbation theory, but exhibits dynamical mass generation 
non-perturbatively and a mass gap in the spectrum. It is asymptotically 
free, as is QCD, and $m$, the dynamical mass of the 
$\sigma$-particle, is the analogue of the QCD scale $\Lambda$. We 
cannot go into the details of the model here except to say that 
the vacuum expectation value of the square of the auxiliary 
field $\alpha(x)$,
\begin{equation}
\label{alphasquared}
\langle \alpha^2\rangle(\mu,m) = \int\limits_{p^2<\mu^2} 
\frac{d^2p}{(2\pi)^2}\,4\pi\sqrt{p^2 (p^2+4 m^2)}\left[\ln\frac{
\sqrt{p^2+4 m^2}+\sqrt{p^2}}{\sqrt{p^2+4 m^2}-
\sqrt{p^2}}\right]^{-1}\!,
\end{equation}
can be considered as the $\sigma$-model 
analogue of $\langle \alpha_s G^2\rangle$. 
Note that the restriction $p^2<\mu^2$ defines the otherwise 
singular operator product $\alpha^2$. The integral can be 
evaluated\,\cite{NSVZ84,physrep} with the result
\begin{equation}
\label{vev1}
\langle \alpha^2\rangle(\mu,m) = m^4\left[\mbox{Ei}(\ln A)+
\mbox{Ei}(-\ln A)-\ln\ln A -\ln(-\ln A) - 2\gamma_E\right],
\end{equation}
where $\gamma_E=0.5772\ldots$ is Euler's constant, $\mbox{Ei}(-x)=
-\int_{x}^{\infty} dt\, e^{-t}/t$ the exponential integral function and 
\begin{equation}
A = \left(\sqrt{1+\frac{\mu^2}{4 m^2}}+\sqrt{\frac{\mu^2}{4 m^2}}
\,\right)^4.
\end{equation}
Note that $F(x)\equiv
\mbox{Ei}(-x)-\ln x$ has an essential singularity at $x=0$ 
but no discontinuity. 

Equation~(\ref{vev1}) is proportional to the dynamically generated scale 
to the fourth power, as expected, but for $\mu\gg m$ it develops 
power-like cutoff dependence. To see the emergence of renormalons, 
we expand the vacuum expectation value in powers of 
$m/\mu$ and the $\sigma$-model coupling 
$\hat{g}(\mu)\equiv-\beta_0 g(\mu) = 1/\ln(\mu^2/m^2)$. To perform 
the expansion we need the asymptotic expansion 
of $F(x)$ at large $x$. For positive argument the asymptotic expansion 
is
\begin{equation}
F(x) = -\ln x + e^{-x} \sum_{n=0}^\infty (-1)^{n+1}\,\frac{n!}{x^{n+1}}.
\end{equation}
If the divergent series is understood as its Borel sum, the right hand 
side equals $F$. For negative, real argument, one obtains the 
asymptotic expansion
\begin{equation}
\label{aspos}
F(-x) = e^x \left[\,\sum_{n=0}^\infty \frac{n!}{x^{n+1}} - e^{-x}
\left(\ln x \mp i\pi\right)\right].
\end{equation}
Note the ``ambiguous'' imaginary part in the exponentially small term. 
The interpretation of 
Eq.~(\ref{aspos}) is as follows: the upper (lower) sign is 
to be taken, if the (non-Borel-summable!) 
divergent series is interpreted as the Borel 
integral in the upper (lower) complex plane. With this interpretation   
Eq.~(\ref{aspos}) is exact and unambiguous. Inserting these expansions, 
the condensate is given by
\begin{eqnarray}
\label{vev2}
\langle \alpha^2\rangle(\mu,m) &=& \mu^4\sum_{n=0}^\infty \left(
\frac{\hat{g}(\mu)}{2}\right)^{n+1} \!\!\!\!n! \,+ \,2\hat{g}(\mu)\,\mu^2 m^2 
\nonumber\\ 
&&\hspace*{-1.5cm} + \,m^4
\left[-2\ln\frac{2}{\hat{g}(\mu)} \pm i\pi-2\gamma_E-4\hat{g}(\mu)+
\frac{\hat{g}(\mu)^2}{2}\right] + O\!\left(\frac{m^6}{\mu^2}\right).
\end{eqnarray}
The expansion for large $\mu$ has quartic and quadratic terms in 
$\mu$, parametrically larger than the ``natural magnitude'' of the 
condensate of order $m^4$. The power terms in $\mu$ arise from the 
quartic and quadratic divergence of the Feynman integral 
(\ref{alphasquared}), i.e. from loop momentum $p\sim \mu$. The 
$\mu$-dependence cancels with the $\mu$-dependence of the coefficient 
functions in the OPE. In particular the $\mu^4$-term cancels with the 
coefficient function of the unit operator. The important point to note 
is that the condensate is unambiguous, but separating the 
``perturbative part'' of order $\mu^4$ is not, since the asymptotic 
expansion for $\mu/m\gg 1$ leads to divergent, non-sign-alternating 
series expansions, which require a summation prescription. The 
``non-perturbative part'' of order $m^4$ depends on this prescription 
(via $\pm i\pi$ in Eq.~(\ref{vev2})). In a purely perturbative calculation,  
one would only obtain the divergent series expansion. The infrared 
renormalon ambiguity of this expansion would lead us to correctly 
infer the existence of a non-perturbative power correction of  
order $m^4$. For quantities without an OPE this is one of the 
main motivations for considering IR renormalon divergence. 

The renormalon ambiguity does not allow us in general 
to say much about the 
magnitude of the power correction which is determined by other 
terms, such as $\ln(2/\hat{g})$ in Eq.~(\ref{vev2}). The 
$\sigma$-model is somewhat special in this respect, since the 
power-like ambiguities in defining perturbative expansions are 
also parametrically smaller in $1/N$ than the actual condensates. This 
tells us that some caution is necessary in identifying the 
magnitude of the ``renormalon ambiguity'' with the magnitude of power 
corrections. It is probably more appropriate to say that power 
corrections are expected to be at least as large as 
perturbative ambiguities. On the other hand, 
a similar parametric suppression of perturbative ambiguities 
does not seem to take place in QCD.

\subsection{The Large-$\beta_0$ Limit}

The best representative of renormalon divergence is the set of 
fermion ``bubble diagrams''. Renormalons have originally been 
discovered\,\cite{GN74,L77,tH77} in this set of diagrams. It is 
important to bear in mind that the concept of renormalons is 
more general and that all diagrams eventually contribute to 
the overall constant $K$ that appears for example in 
Eq.~(\ref{eq:ansatz}). However, the bubble graphs are useful 
for explicit calculations. With a certain amount of extrapolation  
they have also turned out to give useful approximations to 
perturbative expansions in QCD.

\begin{figure}[t]
   \vspace{-2.4cm}
   \epsfysize=21cm
   \epsfxsize=14cm
   \centerline{\epsffile{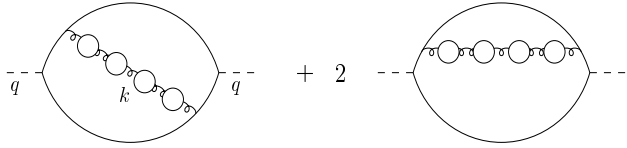}}
   \vspace*{-16.2cm}
\caption[dummy]{\small The set of ``bubble'' diagrams 
consists of all diagrams with any number of fermion 
loops inserted into a single gluon line. \label{fig2}}
\end{figure}

We consider again the current correlation function defined in 
Eq.~(\ref{currentcorr}), more precisely the Adler function 
$D(Q)=4 \pi^2\,d\Pi(Q)/dQ^2$, and compute the ``bubble diagrams'' 
shown in Fig.~\ref{fig2}. Any number of quark loops may be 
inserted into the gluon line; each loop gives a factor 
$\beta_{0f}\alpha_s\left[\ln(-k^2/\mu^2) - 5/3\right]$, if 
the strong coupling is renormalized in the $\overline{\rm MS}$ scheme,
and $\beta_{0f}=-N_f/(6\pi)$ is the quark contribution to the 
1-loop $\beta$-function.

Bubble diagrams can be computed in several ways. 
Originally\,\cite{Ben93a,Bro93} the Borel transform was 
computed directly. The result is\,\cite{Bro93} 
\begin{equation}
\label{borelpolaroper} 
B[D](u)= \frac{32}{3\pi} 
\left(\frac{Q^2}{\mu^2} e^{-5/3}\right)^{-u} \frac{u}{1-(1-u)^2} \,
\sum_{k=2}^\infty \frac{(-1)^k k}{(k^2-(1-u)^2)^2}
\end{equation}
with $u=\beta_{0f} t$. This has the singularities shown in
Fig.~\ref{fig1} except for the presence of $\beta_{0f}$ rather 
than $\beta_0$. The origin of the singularities is more evident, 
if we integrate over the loop momentum 
of the ``large'' quark loop in Fig.~\ref{fig2} 
and the angles of the gluon momentum 
$k$. Defining $l=-k^2/Q^2$, we obtain
\begin{equation}
\label{basint}
D = \sum_{n=0}^\infty \,\alpha_s\int_0^\infty \,\frac{dl}
{l}\,F(l)\,\left[-\beta_{0f}\alpha_s\ln\left(
l\frac{Q^2e^{-5/3}}{\mu^2}\right)\right]^n,
\end{equation}
where\,\cite{N95} 
\begin{eqnarray}
   F(l) &\!=\!& \frac{8 l}{3\pi}\bigg\{
    \bigg( {7\over 4} - \ln l \bigg)\,l
    + (1+l)\,\Big[ \mbox{Li}_2(-l) + \ln l\,\ln(1+l) \Big]
    \bigg\} \quad l<1, \nonumber\\
   && \nonumber\\
   F(l) &\!=\!&  \frac{8l}{3\pi}\bigg\{ 1 + \ln l
    + \bigg( {3\over 4} + {1\over 2}\,\ln l \bigg)\,{1\over l}
    \nonumber\\
   &&\qquad \mbox{}+ (1+l)\,\Big[ \mbox{Li}_2(-l^{-1}) - \ln l\,
    \ln(1+l^{-1}) \Big] \bigg\} \quad l>1 
\end{eqnarray}
with $\mbox{Li}_2(x)$ the dilogarithm. The dominant 
contributions to the integral come from 
$l\gg 1$ and $l\ll 1$, because of the large logarithmic 
enhancements in these regions. There is a one-to-one correspondence 
between each term in the expansion for small (large) $l$ and the 
IR (UV) renormalon poles in the Borel transform. For instance, 
the leading term at small $l$, 
\begin{equation}
\label{smallk}
F(l) = \frac{2}{\pi}\,l^2 + {\cal O}(l^3 
\ln l)
\end{equation}
leads to 
\begin{equation}
\label{dex}
D^{\rm IR} \sim \frac{1}{\pi} \,\frac{\mu^4}{Q^4} \,e^{10/3}
\sum_{n=0}^\infty \alpha_s^{n+1} 
\left(\frac{\beta_{0f}}{2}\right)^n n!,
\end{equation}
which corresponds to the IR renormalon pole at $u=2$.

We can verify explicitly the general result that this expansion 
can be interpreted as part of the gluon condensate. When $k\to 0$ 
the gluon line in Fig.~\ref{fig2} can be ``cut'', i.e. supposed 
to end in a slowly varying external field. The result of the 
computation is $2\pi/(3 Q^4)$, the leading-order 
coefficient function of the gluon condensate. To verify that 
indeed
\begin{equation}
\label{dex2}
D^{\rm IR} \sim \frac{2 \pi}{3 Q^4}  \,
\langle \frac{\alpha_s}{\pi} G^2\rangle(k<\mu),
\end{equation}
we compute the (perturbative) gluon condensate in the bubble 
approximation and obtain\,\cite{Z92,BZ93}
\begin{eqnarray}
\label{basint2}
\langle \frac{\alpha_s}{\pi} G^2\rangle(k<\mu)
&=& \frac{3}{2\pi^3}
\sum_{n=0}^{\mu^2} \,\alpha_s\int\limits_0^\infty dk^2\,k^2
\left[-\beta_{0f}\alpha_s\ln\left(
\frac{k^2e^{-5/3}}{\mu^2}\right)\right]^n
\nonumber\\
&& \hspace*{-2cm}
=\,\frac{3}{2\pi^3} \,\mu^4 e^{10/3}\sum_{n=0}^\infty \alpha_s^{n+1} 
\left(\frac{\beta_{0f}}{2}\right)^n n!.
\end{eqnarray}
Combining this result with the coefficient function, there is 
agreement with Eq.~(\ref{dex}).

The problem with the bubble approximation in QCD is that the
renormalon singularities are determined by the quark contribution 
to the $\beta$-function. The general arguments tell us that 
$\beta_{0f}$ must get converted into $\beta_0$. We could add 
gluon and ghost bubbles as well, but this would still not give a 
complete result. In one way or another, recovering $\beta_0$ 
in QCD leads beyond the approximation of a single dressed gluon 
line. Even if we could construct an effective charge analogous to 
QED, giving the complete $\beta_0$ for every dressed gluon line, 
it is not clear what would be gained from such a construction, since 
the overall normalization of renormalon divergence remains as elusive 
as before, and this constant is all that is not already determined 
by renormalization group arguments.

Despite these diagrammatic difficulties it has been 
suggested\,\cite{BB95a,N95,BBB95,LM95} to use the quark bubble 
calculation with $\beta_{0f}$ replaced by $\beta_0$ as a realistic  
approximation to the full QCD perturbative expansion. Formally, 
this amounts to rewriting a perturbative coefficient at 
order $\alpha_s^{n+1}$ as
\begin{equation}
\label{flavourseries2}
r_n=r_{n0}+r_{n1} N_f+\ldots+r_{nn} N_f^n = r_0
\left[d_n \beta_0^n+\delta_n
\right],
\end{equation}
where $d_n=(-6\pi)^n r_{nn}/r_0$, $\beta_0=(11-2N_f/3)/(4 \pi)$ 
and $N_f$ is the number of massless quarks. The  
coefficients $d_n$ are then obtained 
from a calculation of fermion bubble graphs, while 
the remainder $\delta_n$ is neglected. For this reason, this 
procedure is often referred to as the ``large-$\beta_0$ 
approximation''. It can also be viewed as an extension 
of Brodsky-Lepage-Mackenzie scale setting.\,\cite{BLM83}  
It is more an empirical observation than a well-reasoned statement 
that the remainder is indeed often found to be small compared 
to the $\beta_0$ term (in the $\overline{\rm MS}$ scheme). Since 
the large-$\beta_0$ limit also incorporates the expected divergence 
of the QCD expansion, it has turned out to be a useful quantitative 
tool to estimate higher-order coefficients, in particular when 
the onset of divergence is rapid and when the observable depends 
only on a single scale.

\subsection{IR Renormalons and Power Corrections in 
Quantities without an Operator Product Expansion}

Up to now we have been characterizing infrared renormalon divergence 
using known operator product expansion (OPE) 
methods. Much of the interest in IR renormalons 
during the recent 5 years comes from adopting a somewhat different 
point of view. Rather than concentrating on quantities whose
OPE is well understood, let us take any 
``hard''  quantity, i.e.\ a quantity that admits a perturbative calculation.
If we are able to identify a
particular pattern of IR renormalon divergence, we must replace 
the IR sensitive perturbative contributions by a non-perturbative 
parameter of similar magnitude. In this way, the suppression of 
power corrections can be obtained in a very general way.

To our knowledge this possibility was first mentioned  
by Mueller,\,\cite{Mue85} who suggested to investigate hadronic 
event shape observables and jet cross sections from this perspective. 
The idea fell into oblivion at the time and it took another 
decade before renormalons in observables without an OPE were 
beginning to be studied in detail, first in heavy quark 
physics\,\cite{BB94a,BSUV94} and soon after for event shape 
observables.\,\cite{MW95,Web94} 

It is characteristic of off-shell processes that IR renormalons 
occur only at positive integer multiples of $1/\beta_0$, which 
implies power corrections in powers of $1/Q^2$ and not powers of 
$1/Q$, where $Q$ is the ``hard'' scale of the process. On-shell 
quantities have a large variety of infrared-sensitive regions 
in their Feynman amplitudes and the generic 
situation leads to IR renormalons at positive half-integers and integers 
and a series of power corrections in $1/Q$. Due to the variety of 
possibilities, we restrict ourselves here to a few very general 
remarks. The most interesting cases are then discussed in detail 
in Sect.~\ref{sec:applications}.

The existence of $1/Q$ power corrections in observables related to 
on-shell Green functions can already be seen from the simplest 
example, the two-point function of a heavy quark field, $\Sigma(p,m)$. 
The 1-loop contribution is, schematically,
\begin{equation}
\label{eq:self}
\Sigma(p,m)\propto \int\frac{d^4 k}{(2\pi)^4}
\frac{N}{k^2 (2p\cdot k+k^2+[p^2-m^2])}. 
\end{equation} 
Only the denominator is important for the following discussion. 
Off-shell ($p^2\not=m^2$) the contribution to the integral 
from $k<\Lambda$ is of order $C\Lambda^2/m^2$ relative to the 
dominant contribution from $k\sim m$. As $p^2\to m^2$ the 
coefficient $C$ diverges. Indeed, at $p^2=m^2$ the integrand of 
Eq.~(\ref{eq:self}) vanishes only linearly with $k$ for small $k$; 
hence we expect a non-perturbative contribution to 
$\Sigma(m,m)$ of relative order $\Lambda/m$. 

Similar power counting 
arguments apply to hadronic event shape observables. The basic 
parton emission processes in QCD are infrared divergent for 
soft and collinear emissions. Event shape observables are constructed 
to suppress these soft and collinear contributions, so that 
they are infrared finite in perturbation theory. But the residual 
contribution from small momentum regions is usually suppressed 
only by a single power of the hard scale. In general high-energy
processes involving massless quarks the infrared contributions 
can be classified as soft or hard-collinear. It appears, however, 
that power corrections from hard-collinear regions 
(energy $\omega$ much larger than transverse momentum $k_\perp$) 
are always suppressed by powers of $Q^2$ rather than $Q$. We do not 
know of a proof of this statement, but the following heuristic 
argument may illustrate the point: let $p$ be the momentum of 
a fast on-shell particle, $p\sim Q$, after emission of a hard-collinear 
on-shell particle with $\omega\sim Q$, $k_\perp\ll Q$, where 
$k_\perp$ is the transverse momentum relative to $p$. Then the propagator 
\begin{equation}
\frac{1}{(p+k)^2} = \frac{1}{p\left(\omega-\sqrt{\omega^2+k_\perp^2}
\right)}
\end{equation}
is expanded in $k_\perp^2/\omega^2\sim k_\perp^2/Q^2$ and $Q$ enters 
only quadratically. Since the same is true of the hard-collinear 
phase space, it may be argued that the transverse momentum, and hence $Q$, 
always enters quadratically as long as energies are large.
As a consequence, if $1/Q$ power corrections exist and if one is interested 
only in those, the analysis simplifies, because only soft
contributions need to be considered. 

A systematic analysis of IR renormalons for general short-distance 
processes has therefore much in common with the analysis of 
IR finiteness with the methods of perturbative factorization. 
It extends the notion of IR safety (absence of logarithmic
divergences) to that of IR sensitivity (power-suppressed IR
contributions). From a very general point of view, the most important
lesson drawn from IR renormalons is the existence of a correlation 
between the size of non-perturbative corrections and the 
size of perturbative coefficients in large orders, often already 
at 2 loops.

\subsection{The Landau Pole}

In closing this overview section we address a common misunderstanding 
that the existence of renormalons and summation ambiguities is 
related to an infrared  Landau pole in the running coupling. 
A consequence of this misunderstanding is that it is often thought 
that the power corrections identified via IR renormalons have 
something to do with the definition of the QCD coupling. It is true 
that the factorial divergence follows from the fact that the coupling 
evolves ($\beta_0\not=0$). The power correction indicated by this 
divergence is, however, a property of the particular observable and 
it would exist even if the coupling did not evolve.

To see how the association with the Landau pole might arise, we 
interchange the summation and integration in Eq.~(\ref{basint}). 
The interchange is mathematically not legitimate, but if we
nonetheless proceed, we obtain  
\begin{equation}
\label{drun}
D = \int_0^\infty\,\frac{dl}
{l}\,F(l)\,\alpha_s\!\left(k e^{-5/6}\right),
\end{equation}
where $k^2 = -l Q^2$ and
\begin{equation}
\label{onelooprun}
\alpha_s(k) = \frac{\alpha_s(\mu)}{1+\alpha_s(\mu)\beta_{0}\ln(k^2/\mu^2)} 
\equiv \frac{1}{\beta_{0}\ln k^2/\Lambda^2}
\end{equation}
is the 1-loop running coupling. The Landau pole of the coupling 
lies on the $l$-integration contour and the ambiguity in defining 
the integral (\ref{drun}) due to this Landau pole 
is exactly identical to the ambiguity in 
defining the Borel integral of the divergent series expansion.

This apparent connection does not persist beyond the calculation 
of bubble diagrams. The reason is that in the same approximation in 
which we keep only the bubble diagrams, the $\beta$-function can 
only have a single term: $\beta(\alpha_s)=-\beta_0\alpha_s^2$. The 
existence of a Landau pole is then an automatic consequence. 
The general theory of 
renormalons shows that the leading asymptotic behavior depends 
only on $\beta_0$ and $\beta_1$, see 
Eqs.~(\ref{eq:ansatz},\ref{eq:ab}). On the other 
hand whether a Landau pole 
exists or not is a strong-coupling problem and it depends on all  
coefficients of the $\beta$-function, and on 
power corrections to the running of the coupling. This can 
be studied explicitly for $\beta$-functions with only two 
independent terms.\,\cite{Gru96,DU96,PdR96} For example, the 
integral (\ref{drun}) might be well-defined, but its series 
expansion remains still divergent. It would then be incorrect 
to conclude that the power corrections indicated by IR renormalons 
are properly taken into account by summing the series to the numerical
value given by the integral. Such a summation prescription 
would be related to a certain definition of non-perturbative 
parameters relevant to the particular process, but it would not 
render these parameters zero and superfluous. In essence, 
IR renormalons reflect perturbative aspects of non-perturbative 
corrections (namely their power divergences) for a specific 
observable, whereas the existence 
of a Landau pole is a wholly non-perturbative issue, but not related to 
any particular observable. This can again be non-perturbatively 
verified in the non-linear $O(N)$ $\sigma$-model,\,\cite{BBK98} where 
the propagator of the $\alpha$-field defines an effective coupling 
without Landau pole, but all correlation functions contain the full 
series of IR renormalon poles.

\section{Applications}
\label{sec:applications}

\subsection{Deep Inelastic Scattering}

We begin with a short exposition of the applications of renormalons
to the structure functions of deep inelastic scattering.
Since the operator product expansion is available, this example
serves to illustrate the operator interpretation of renormalons 
and formulate phenomenological procedures that can be generalized 
to other, more complicated situations.  

\subsubsection{The GLS Sum Rule}

The GLS sum rule, defined in Eq.~(\ref{eq:gls}), 
has already been discussed in the introduction; here we complete 
this discussion.

The first IR renormalon singularity at $t=1/\beta_0$ corresponds to 
a single twist-4 operator\,\cite{Mue93} and can be obtained with 
the methods described in Sect.~\ref{sec:basics}. Combining this 
result with the calculation of the leading ultraviolet 
renormalon\,\cite{BBK97} (see Sect.~\ref{sec:uvrenormalon}), we find 
the large-order behavior
\begin{equation}
r_n \sim
 \beta_0^n\,n!\,\Big[K^{\rm UV}\,(-1)^n n^{1-\beta_1/\beta_0^2+\lambda_1} 
+\,K^{\rm IR}\,n^{\beta_1/\beta_0^2-32 b/9}
\Big],
\label{dis2}
\end{equation}
where $r_n$ is the coefficient of $\alpha_s^{n+1}$, 
$\beta_{0,1}$ are the first two coefficients of the 
$\beta$-function, $b = 4\pi\beta_0 = 11-2 N_f/3$ and
$\lambda_1$ is related to the anomalous dimension matrix of four-fermion 
operators, see Sect.~4. For $N_f>2$, the UV renormalon behavior 
dominates the asymptotic behavior at very large $n$ because of its larger 
power of $n$. However, the overall normalization constants 
$K^{\rm UV}$ and $K^{\rm IR}$ are not known. 
Since the $\overline{\rm MS}$ scheme favors large residues of IR 
renormalons, at least in the large-$\beta_0$ approximation, we expect 
fixed-sign IR renormalon behavior in intermediate orders. The first
three terms in the series known exactly are indeed of the same sign 
in the $\overline{\rm MS}$ scheme as can be seen from 
Eq.~(\ref{eq:gls}).

Is the asymptotic behavior in Eq.~(\ref{dis2}) relevant to  
phenomenology? Since the constants $K$ are not known, we 
consider\,\cite{Ji95,LM95} the large-$\beta_0$ approximation.
The Borel transform (defined by Eq.~(\ref{defborelt})) of 
the perturbative expansion in this approximation 
is given by\,\cite{BK93}
\begin{equation}
\label{borelgls}
B[1-I_{\rm GLS}/3](u) = 
\left(\frac{Q^2}{\mu^2}\,e^C\right)^{-u}\,\frac{1}{9\pi}\left\{
\frac{8}{1\!-\!u}+\frac{4}{1\!+\!u}-
 \frac{5}{2\!-\!u}-\frac{1}{2\!+\!u}\right\},
\end{equation}
where $u=\beta_0 t$. This is much simpler than what we would have
expected  
on general grounds. In particular, there are only four 
renormalon poles, all 
others being suppressed in the large-$\beta_0$ limit. The structure of 
the leading singularities is also simpler than the exact result 
in Eq.~(\ref{dis2}),  because the anomalous dimensions should be set to 
zero in this limit. Nonetheless, the exact $r_1,r_2$ are reproduced 
reasonably well by the large-$\beta_0$ limit. 
The large-$\beta_0$ approximation taken at 
face value implies that the 
minimal term of the series is reached at order $\alpha_s^{3}$
or $\alpha_s^{4}$ 
at $Q^2=3\,$GeV${}^2$, a momentum transfer relevant to the CCFR 
experiment. Hence it is not clear whether at $Q^2=3\,$GeV${}^2$ 
the perturbative prediction could be improved further by exact 
calculations of higher-order corrections. Further improvement would 
then require the inclusion of twist-4 contributions, and in particular 
a practically realizable procedure to combine them consistently 
with the perturbative series. This hypothesis is further supported by 
noting that the integral over 
loop momentum is dominated by $k\sim 450\,$MeV at order $\alpha_s^3$ and 
$k\sim 330\,$MeV at order $\alpha_s^4$.  
The ambiguity in summing the perturbative 
expansion is of order
(assuming $\Lambda_{\overline{\rm MS}}=215\,$MeV)
\begin{equation}
\frac{1}{3}\,\delta I_{\rm GLS} = \frac{1}{\beta_0}
\frac{8 e^{5/3}}{9\pi}\,\frac{\Lambda^2_{\overline{\rm MS}}}
{Q^2} \approx \frac{0.1\,\mbox{GeV}^2}{Q^2}.
\end{equation}
This should be compared to the twist-4 contribution to the 
same quantity estimated 
in quark model and by QCD sum rules\,\cite{BK87} 
\begin{equation}
-\frac{8}{27}\,\frac{\langle\langle {\cal O}_4\rangle\rangle}{Q^2} 
\approx -\frac{0.1\,\mbox{GeV}^2}{Q^2},
\end{equation}
where $\langle\langle {\cal O}_4\rangle\rangle$ is the 
reduced nucleon matrix element of the twist-4 operator. 
The two are comparable, which suggests that the treatment of 
perturbative corrections beyond those known exactly is 
as important for a determination of $\alpha_s$ from the 
GLS sum rule as the twist-4 
correction.

\subsubsection[The $x$-Dependence of 
Power Corrections]{The $x$-Dependence of 
Power Corrections to Structure Functions}

The operator product expansion allows us 
to express $1/Q^2$ corrections
to the structure functions in terms of contributions of several towers
of twist-4 operators\,\cite{JS81} or, equivalently, several multiparton 
correlation functions.\,\cite{EFP83,J83,BB88} Unfortunately, the 
structure of the corrections is complicated and they involve many 
non-perturbative parameters which cannot all be extracted from 
inclusive measurements. Because of this, it has never been possible to use
this sophisticated machinery in the analysis of real data.
In practice, higher-twist corrections are being extracted from data
from a combined fit to perturbative and $1/Q^2$-suppressed
contributions in a large $Q^2$ range.

Renormalons provide a simple ansatz for the $x$-dependence 
of power corrections involving fewer (if any) non-perturbative parameters.
As in most other applications divergences of the perturbative 
series are relevant inasmuch as they originate from small 
momentum regions in Feynman diagrams that are also responsible for  
non-perturbative effects. To explain the idea, recall that the 
structure functions (take $F_2(x)$ as an example) are related to 
the parton  distribution functions $P(x)$, $P = q, \bar q, g$ through 
a perturbative expansion and convolution of the form 
\begin{eqnarray}
F_2(x,Q^2) &=& 2x \sum_{P}\left[e^2_P+\sum_{n=0}r_{P,n}(x)\alpha_s^{n+1}
- c_P(x)\frac{\mu^2}{Q^2}\right]\star P(x,Q^2)
\nonumber\\
&&{} + \frac{C_2(x)}{Q^2},  
\label{opef2}   
\end{eqnarray}   
where $e_P$ are the electromagnetic charges of partons and 
$\star$ stands for the convolution. In this expression we have subtracted 
contributions of small momenta $k < \mu \ll Q$ from the coefficient functions
and defined the higher-twist contribution which can have a complicated 
operator content as the full contribution (perturbative and non-perturbative)
coming from small momenta $k < \mu$. Since twist-4 operators are quadratically
divergent, we know that $C_2(x) = c'(x)\mu^2 + O(\Lambda^2)$ if 
$\mu\gg \Lambda$ and $c'(x) = \sum_P c_P(x)\star P(x)$ 
so that the dependence on the cutoff $\mu$ cancels.

As $c(x)$ is defined as the small-momentum contribution 
to Feynman diagrams, it has itself a (divergent) perturbative expansion 
in $\alpha_s$. We then extract the $c(x)$ from the $x$-dependence of 
the IR renormalon pole computed in the large-$\beta_0$ approximation. 
In this approximation there is only a (anti-)quark contribution and    
the result is\,\cite{BB95b,DMW96,SMMS96,DW96a}
\begin{eqnarray}
c_q^{(L)}(x) &=& 8 x^2-4\delta(1-x), 
\\
c_q^{(2)}(x) &=& -\frac{4}{[1-x]_+}+4+2 x+12 x^2-9\delta(1-x)-\delta'(1-x), 
\\
c_q^{(3)}(x) &=& -\frac{4}{[1-x]_+}+4+2 x+ 4 x^2-5\delta(1-x)-\delta'(1-x)
\label{c2dis}
\end{eqnarray}
for $F_L$, $F_2$ and $F_3$, respectively. A common 
overall normalization is omitted here, because it plays no role 
in the following. 
The `+' prescription is defined as usual by 
$\int_0^1d x \,[f(x)]_+ t(x) = \int_0^1 d x \,f(x) \,(t(x)-t(1))$ 
for test functions $t(x)$. 

If we assume that the subtraction term in the square brackets of 
Eq.~(\ref{opef2}) cancels approximately the higher-order 
perturbative terms, and if we assume that $C_2(x)$ is 
approximated by $c'(x)\mu^2$, we obtain an improved prediction 
for the structure functions compared to the fixed-order 
perturbative approximation. This suggestion, though sometimes
motivated by different arguments, has become known as ``renormalon model'' 
of power corrections.\,\cite{DMW96,SMMS96} 
The structure functions are then written as
\begin{equation}
\label{par1}
F(x,Q) = F^{\rm tw-2}(x,Q)\left(1+\frac{D(x,Q)}{Q^2}+ O(1/Q^4)
\right),
\end{equation}
where $F^{\rm tw-2}(x,Q)$ is the leading-twist result for the 
structure function $F = F_L, F_2, F_3,\ldots$ and 
\begin{equation}
\label{par}
D(x,Q) = \frac{ \Lambda^2 }{F^{\rm tw-2}(x,Q)}
 \int_x^1\frac{d\xi}{\xi}\,c(\xi)\,q(x/\xi,\mu)\, 
\end{equation}
is the model parametrization of the (relative) twist-4 correction. 
Here $q(x,\mu)$ is the  standard (leading-twist) quark density, 
and $\Lambda$ is a certain scale of 
order $\Lambda_{\rm QCD}$ which provides the overall normalization. 
The expression can be extended to include gluon contributions and/or
higher-order corrections should they become available.

The overall normalization has been treated differently in the 
literature. One suggestion has been\,\cite{DMW96} to parametrize 
the normalization of all $1/Q^2$ power corrections by a single 
process-independent number, to be extracted from the data once
and related to a a certain effective QCD coupling. 
Other authors\,\cite{MSSM97,BBM97,YB99} prefer to adjust the 
normalization in a process-dependent way and to take only the 
shape of the $x$-distribution as a prediction of the model. 
Because of difficulties in constructing the gluon contribution 
in the model, one may also think of adjusting the normalization of 
quark and gluon contributions separately. 

\begin{figure}[t]
   \vspace{0.0cm}
   \epsfysize=6.5cm
   \epsfxsize=8.5cm
   \centerline{\epsffile{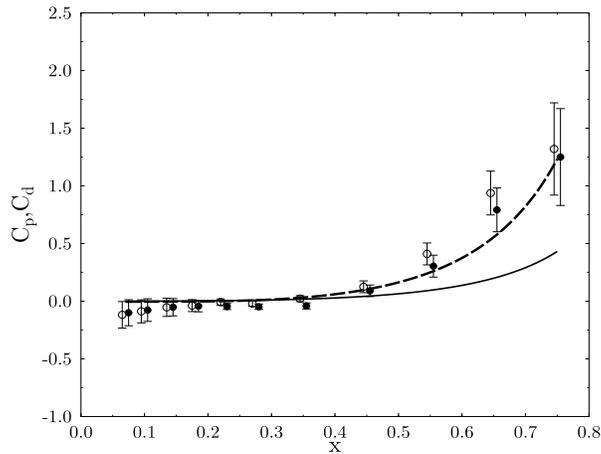}}
   \vspace*{-0.3cm}
\caption[dummy]{\small  Relative twist-4 contribution $D(x)$ 
(called $C_{p,d}(x)$ here)  defined by Eq.~(\ref{par}) 
to the proton (deuteron) structure function $F_2$ in the 
``renormalon model''\,\cite{MSSM97} 
(dashed line) compared with 
proton (filled circles) and deuteron (empty circles) data.\,\cite{VM92} 
The solid curve shows the unrescaled estimate of the renormalon ambiguity.
\label{fig16}}
\end{figure}
The ``renormalon model'' of twist-4 corrections 
has first been applied 
to the structure function $F_2$.\,\cite{DMW96,DW96a,MSSM97}  
As shown in Fig.~\ref{fig16}, the 
shape of the twist-4 correction calculated from the model indeed 
reproduces the experimental data very well. These 
results refer to the the non-singlet contribution 
to $F_2$, which is expected to dominate except for small values of $x$. 
Similar predictions have been worked out for
the longitudinal structure function $F_L$,\,\cite{SMMS96,DW96a}
$F_3$,\,\cite{DW96a,MSSM97} and the polarized structure function 
$g_1$ etc.\,\cite{DW96a,MMMSS96}
More recently a renormalon model prediction has also been 
constructed for the singlet contribution\,\cite{SMMS98,Smy98} 
to $F_2$, which modifies the analysis at small $x$, below those $x$ 
for which comparison with present data is possible.
(The treatment of 
singlet contributions is more difficult and ambiguous in the renormalon model 
than non-singlet contributions. 
The calculation relies on singlet quark contributions, 
which are then reinterpreted as gluon contributions.\,\cite{BBM97})

\begin{figure}[p]
   \vspace{-1.6cm}
   \epsfysize=8cm
   \epsfxsize=5cm
   \centerline{\epsffile{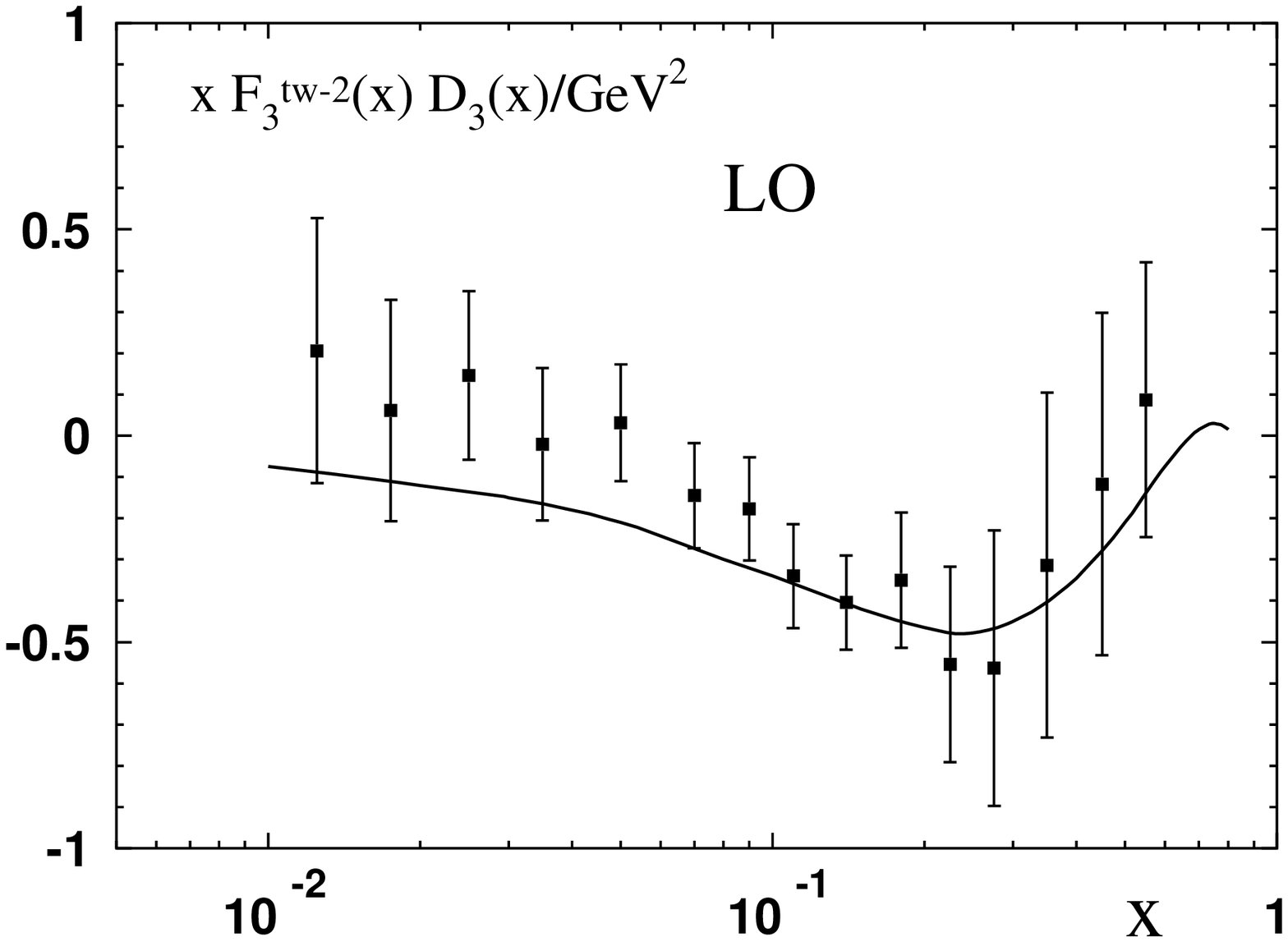}}
   \vspace*{-3.3cm}
   \epsfysize=8cm
   \epsfxsize=5cm
   \centerline{\epsffile{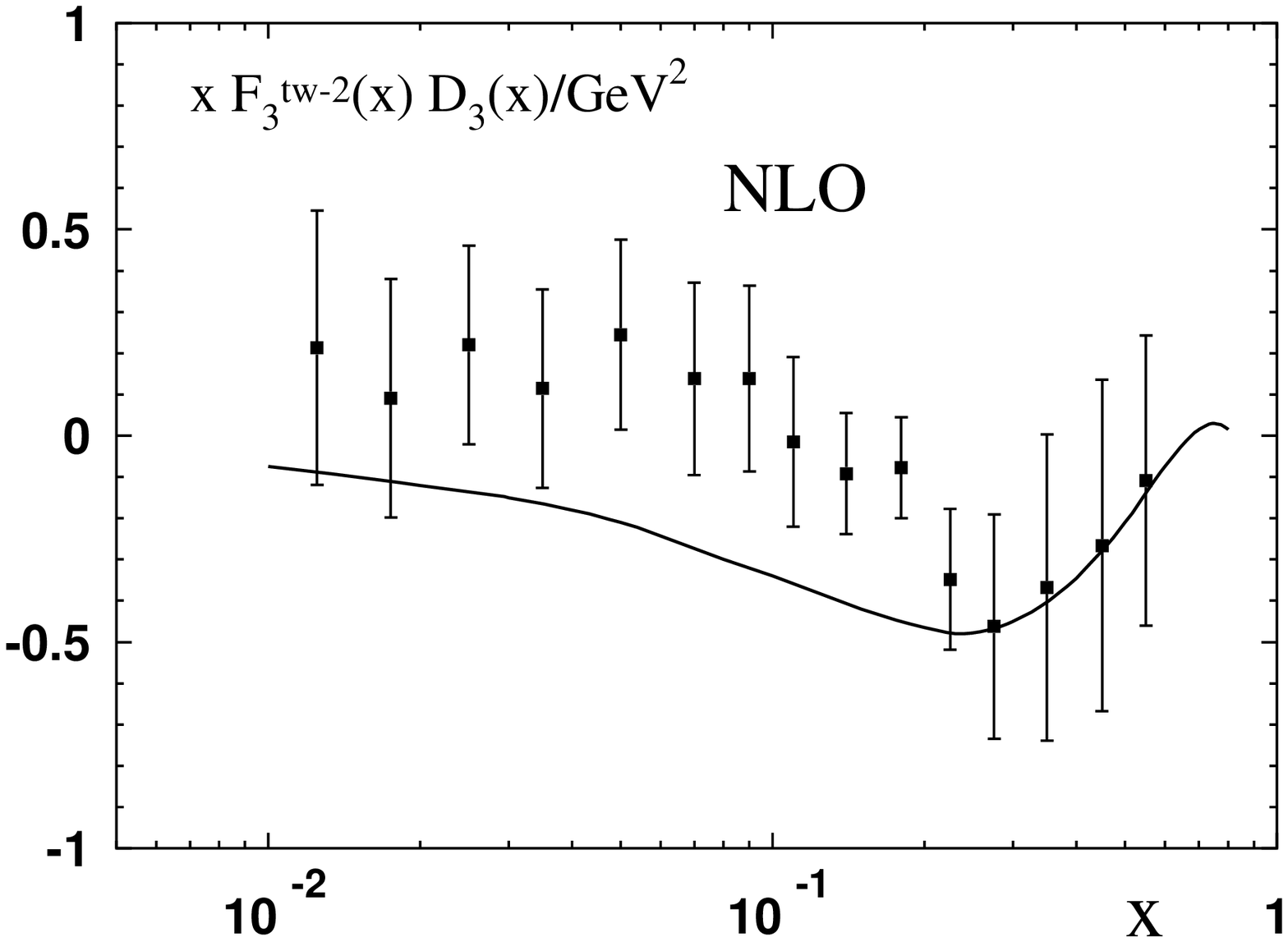}}
   \vspace*{-3.3cm}
   \epsfysize=8cm
   \epsfxsize=5cm
   \centerline{\epsffile{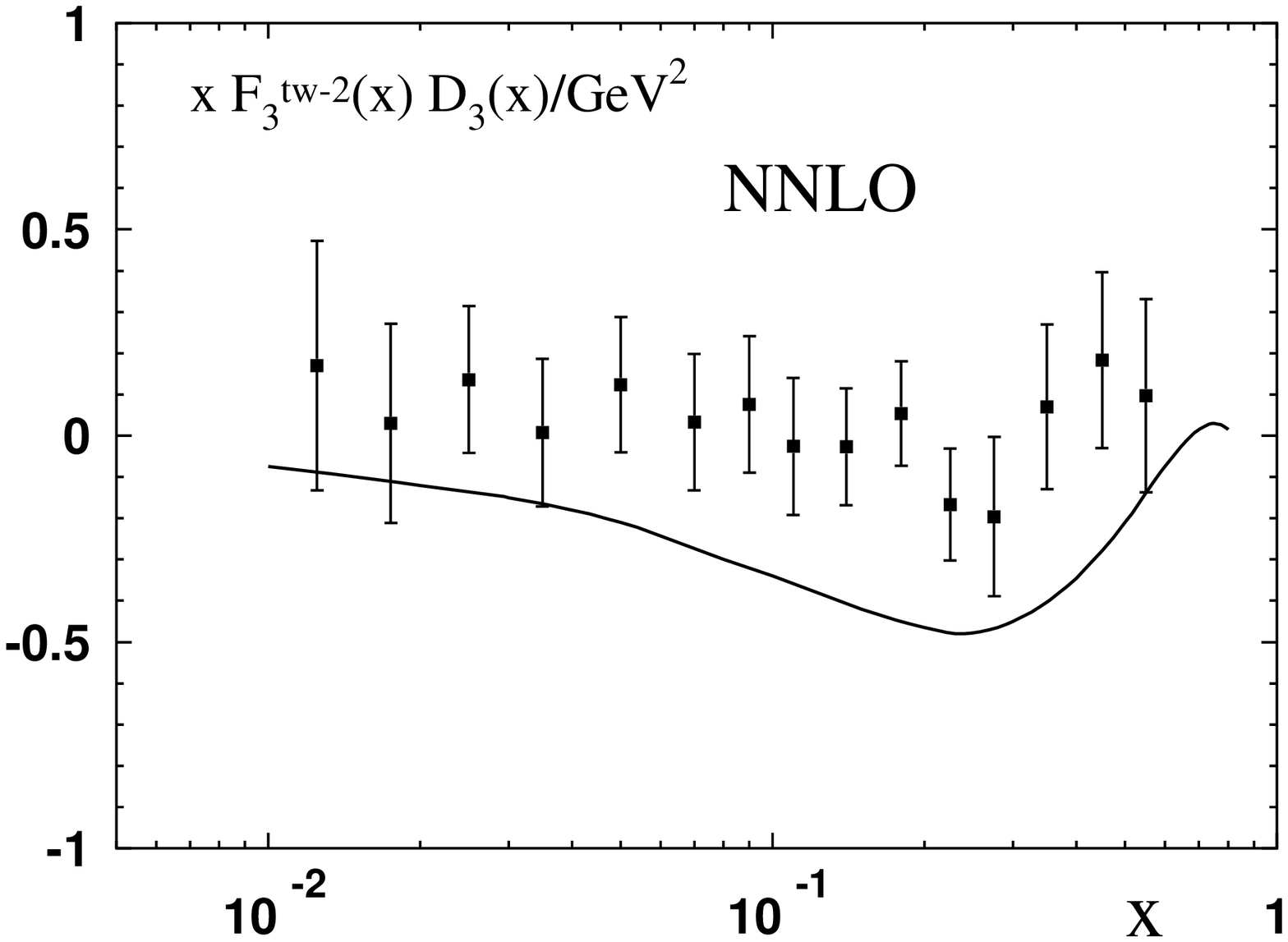}}
   \vspace*{-2cm}
\caption[dummy]{\small Twist-4 correction to $x F_3$ as extracted from 
the (revised) CCFR data. The three plots show the effect of including 
leading order (LO), next-to-leading order (NLO) and 
next-to-next-to-leading order (NNLO) QCD corrections in the 
twist-2 term. The data points\,\cite{KKPS97} are  overlaid with the 
 shape obtained from the 
``renormalon model'' for the $1/Q^2$ power correction. \label{fig17}}
\end{figure}

A striking property of the renormalon model for twist-4 corrections  
is that all target-dependence enters trivially through the 
target dependence of the twist-2 distribution functions. Therefore 
the  renormalon model can be useful only if the 
genuine twist-4 target dependence is small compared to the magnitude of the 
twist-4 correction itself. In terms of 
moments $M_n$, Eq.~(\ref{par}) implies
\begin{equation}
\frac{M_n^{\rm tw-4}}{M_n^{\rm tw-2}}_{|\rm hadron 1} = \,\,\,
\frac{M_n^{\rm tw-4}}{M_n^{\rm tw-2}}_{|\rm hadron 2}\,. 
\end{equation}
Figure~\ref{fig16} shows that this 
is indeed the case for $F_2$ of protons against deuterons, in particular 
in the region of large $x$. More recent analyses\,\cite{YB99,YB00} 
also confirm this fact. 

It is known that higher-twist corrections (as well as higher-order 
perturbative corrections) are enhanced as $x\to 1$.\,\cite{BBL89} 
This is in part an effect of kinematic 
restrictions near the exclusive region and the renormalon model 
reproduces such enhancements.  For the structure functions it is found that 
power corrections related to renormalons are of order 
\begin{equation}
\label{scaledis}
\left[\frac{\Lambda^2}{Q^2 (1-x)}\right]^{\!n},
\end{equation}
at least those related to diagrams with a single gluon line.\,\cite{BB95b}
This tells us that the increase of the 
twist-4 correction towards larger $x$ seen in the model and the data 
in Fig.~\ref{fig16} may to a large extent be 
the correct parametrization of such a kinematic effect. Note 
that Eq.~(\ref{scaledis}) can be understood from the fact that 
the hard scale in DIS is $Q\sqrt{1-x}$ at large 
(but not too large) $x$. 

It is also possible that both the experimental parametrization of 
higher-twist corrections and the model provide effectively a 
parametrization of higher-order perturbative corrections to 
twist-2 coefficient functions. As far as data are concerned, it should be  
kept in mind that it is obtained from subtracting from the measurement 
a twist-2 contribution obtained from a truncated perturbative 
expansion. As far as the renormalon model is concerned, it is best 
justified by the ``ultraviolet dominance hypothesis''.\,\cite{BBM97} 
Since UV contributions to twist-4 
contributions can also be interpreted as contributions to twist-2 
coefficient functions, a ``perturbative'' interpretation of the 
model prediction suggests itself as we have already indicated 
in the discussion of Eq.~(\ref{opef2}). Note that higher-order corrections 
in $\alpha_s(Q)$ vary more rapidly with $Q$ than lower order ones, 
and may not be easily distinguished from a $1/Q^2$ behavior, 
if the $Q^2$-coverage of the data is not very large. An interesting 
hint in this direction is provided by the analysis\,\cite{KKPS97,KPS00,KS00}
 of CCFR data on $F_3$, reproduced in Fig.~\ref{fig17}. 
 The figure shows how the experimentally fitted twist-4 correction 
gradually disappears as NLO and NNLO perturbative corrections 
to the twist-2 coefficient functions are 
included. At the same time, the renormalon model for the 
twist-4 corrections reproduces well
the shape of data at leading order, and hence parametrizes successfully 
the effect of NLO and (approximate) 
NNLO corrections. This is an important 
piece of information, relevant to quantities for which an NNLO or 
even NLO analysis is not yet available.

Note that whether the model is interpreted as a model for twist-4 
corrections or higher-order perturbative corrections is insignificant 
inasmuch 
as renormalons are precisely related to the fact that 
the two cannot be separated unambiguously. The model clearly cannot be 
expected to reproduce fine structures of twist-4 corrections. Its  
appeal draws from the fact that it provides a simple way to 
incorporate some contributions beyond LO or NLO in perturbation 
theory, which may be the dominant source of discrepancy with data 
with the presently achievable accuracy.

\subsection{Hadronic Event Shape Variables}

The structure of hadronic 
final states in $e^+e^-$ annihilation and deep inelastic 
scattering is the subject of intensive ongoing
studies.\,\cite{eshapes}  
This structure is characterized by a set of infrared and 
collinear safe event shape variables that are calculated in perturbative 
QCD in terms of quark and gluon momenta and compared to the measured
hadron distributions. Apart from a correction for detector effects, 
the comparison of theory and data therefore requires a correction for 
hadronization effects that are most commonly modeled using  
Monte Carlo event generators. It has been 
known for quite some time\,\cite{Bar86}
 that the hadronization corrections are substantial.
In this section we review recent developments that 
relate hadronization corrections to power corrections of order 
$\Lambda/Q$ (where $Q$ is the center-of-mass energy in $e^+ e^-$ 
annihilation) indicated by renormalons 
in the perturbative prediction for the event shape variables. 
This connection was suggested\,\cite{Mue85,MW95} and 
worked out for a few cases of practical interest\,\cite{Web94} several
years ago.
These studies provided the first theoretical indications that hadronization
corrections to most of the observables should scale as $\Lambda/Q$, i.e.
are suppressed by only a single power of the large momentum. Subsequent
analyses\,\cite{DW95,AZ95a,NS95,KS95a} confirmed this conclusion.

Most of the discussion below assumes a generic 
event shape observable $S$ that is of order $\alpha_s(Q)$ at 
leading order. One example is $S=1-T$ where ``thrust'' $T$
is defined as
\begin{equation}
\label{thr}
T = \max_{\,\vec{n}} \frac{\sum_i |\vec{p}_i\cdot\vec{n}\,|}{
\sum_i |\vec{p}_i|},
\end{equation}
where the sum is over all hadrons (partons) in the event. The thrust 
axis $\vec{n}_T$ is the direction at which the maximum is attained. 
Other event shapes considered in connection with power corrections 
are the heavy jet mass, jet broadening,
C-parameter and the longitudinal cross section in 
$e^+e^-$ annihilation, to give only a few examples.

\subsubsection{Mean Values of Event Shape Variables}

It is relatively easy to understand that event shape observables 
in $e^+e^-$ annihilation  are 
linearly sensitive to small parton momenta and hence are expected 
to receive non-perturbative contributions of order $\Lambda/Q$.
Consider an event shape variable that is zero at tree level and therefore 
related to the matrix element for gluon emission $\gamma^*\to q\bar q g$
to leading order
\begin{equation}
\langle S \rangle = \int\!\mbox{dPS}[p_i]\,|{\cal M}_{q\bar{q} g}|^2\,
S(p_i).
\label{<S>}
\end{equation} 
It can be argued that the $\Lambda/Q$ sensitivity arises neither 
from emission of collinear and energetic partons, nor from soft quarks,
but only from soft gluons.
Introducing the energy fractions $x_i=2 p_i\cdot q/q^2$, and reserving 
$x_3$ for the gluon energy fraction, this implies that the only relevant 
integration region is $x_3\to 0$ and, therefore, the gluon emission 
can be calculated in the ``soft'' approximation:
\begin{equation}
|{\cal M}_{q\bar{q} g}|^2 = 32 g_s^2\,\frac{2}{(1-x_1)\,(1-x_2)}.
\label{softM}
\end{equation}
The phase space is, on the other hand
\begin{equation}
\int\!\mbox{dPS}[p_i] = \int_0^1 dx_1 dx_2\,\theta(x_1+x_2-1)\,
\end{equation}
and since the matrix element in Eq.~(\ref{softM}) is singular as 
$x_{1,2}\to 1$, there is a potential logarithmic singularity. To obtain an 
IR finite result, the event shape variable has to be constructed so as to 
eliminate this divergence. The generic situation with event shapes 
is a {\em linear} suppression of soft gluons, $S(x_i) \propto x_3$ as 
$x_3\to 0$, e.g.\ for the thrust $1-T = 1 - {\rm max}(x_1,x_2)$ with 
$x_1+x_2+x_3=2$. 
It is easy to see that this property implies 
a contribution of order $\mu/Q$ to the integral in Eq.~(\ref{<S>})
from gluons with energy less than $\mu$, unless there is
some cancelation. 

It has been suggested\,\cite{DW95,AZ95a,KS95b} that the 
leading power correction 
to average event shape observables may be described by a single 
(``universal'') parameter multiplied by an observable-dependent, but 
calculable, coefficient. 
Write 
\begin{equation}
\label{sdwaz}
\langle S \rangle = A_S \alpha_s(\mu) + \left[B_S+A_S\beta_0
\ln\frac{\mu^2}{Q^2} \right]\alpha_s(\mu)^2 +\ldots + 
\frac{K_S(\mu)}{Q} + O(1/Q^2).
\end{equation}
Dokshitzer and Webber\,\cite{DW95} 
para\-met\-rize the coefficient of the power correction 
in the form 
\begin{equation}
\label{ksmu}
K_S(\mu) = \frac{16 c_S}{3\pi} \mu_I\left[\bar{\alpha}_0(\mu_I)-
\alpha_s(\mu)-\left(\beta_0\ln\frac{\mu^2}{\mu_I^2} + \frac{K}{2\pi} +
2\beta_0\right)\alpha_s(\mu)^2\right],
\end{equation}
where $\mu_I$ is an IR subtraction scale (typically chosen to be $2\,$GeV), 
$\bar{\alpha}_0(\mu_I)$ is the non\--per\-tur\-ba\-tive parameter to be 
fitted and $K=67/6-\pi^2/2-5 N_f/9$. 
The remaining terms approximately subtract the IR contributions 
contained in the perturbative coefficients $A$ and $B$ up to second order. 
The universality assumption can be tested by fitting the value of 
$\bar{\alpha}_0(\mu_I)$ or, equivalently, $K_S(\mu)/c_S$ to different 
event shape variables.  

\begin{figure}[p]
   \vspace{-2cm}
   \epsfysize=11cm
   \epsfxsize=7.6cm
   \centerline{\epsffile{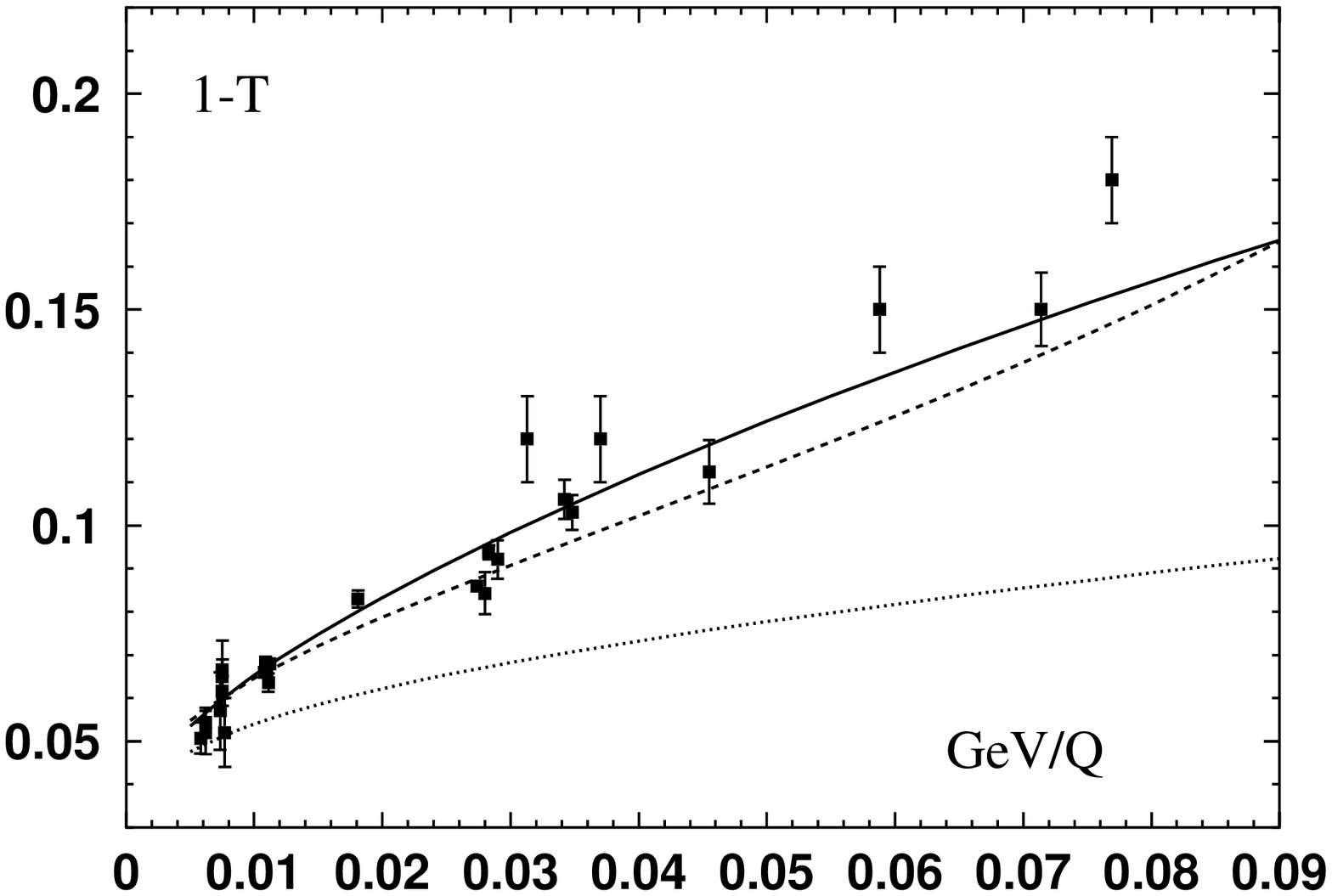}}
   \vspace*{-4.5cm}
   \epsfysize=11cm
   \epsfxsize=7.6cm
   \centerline{\epsffile{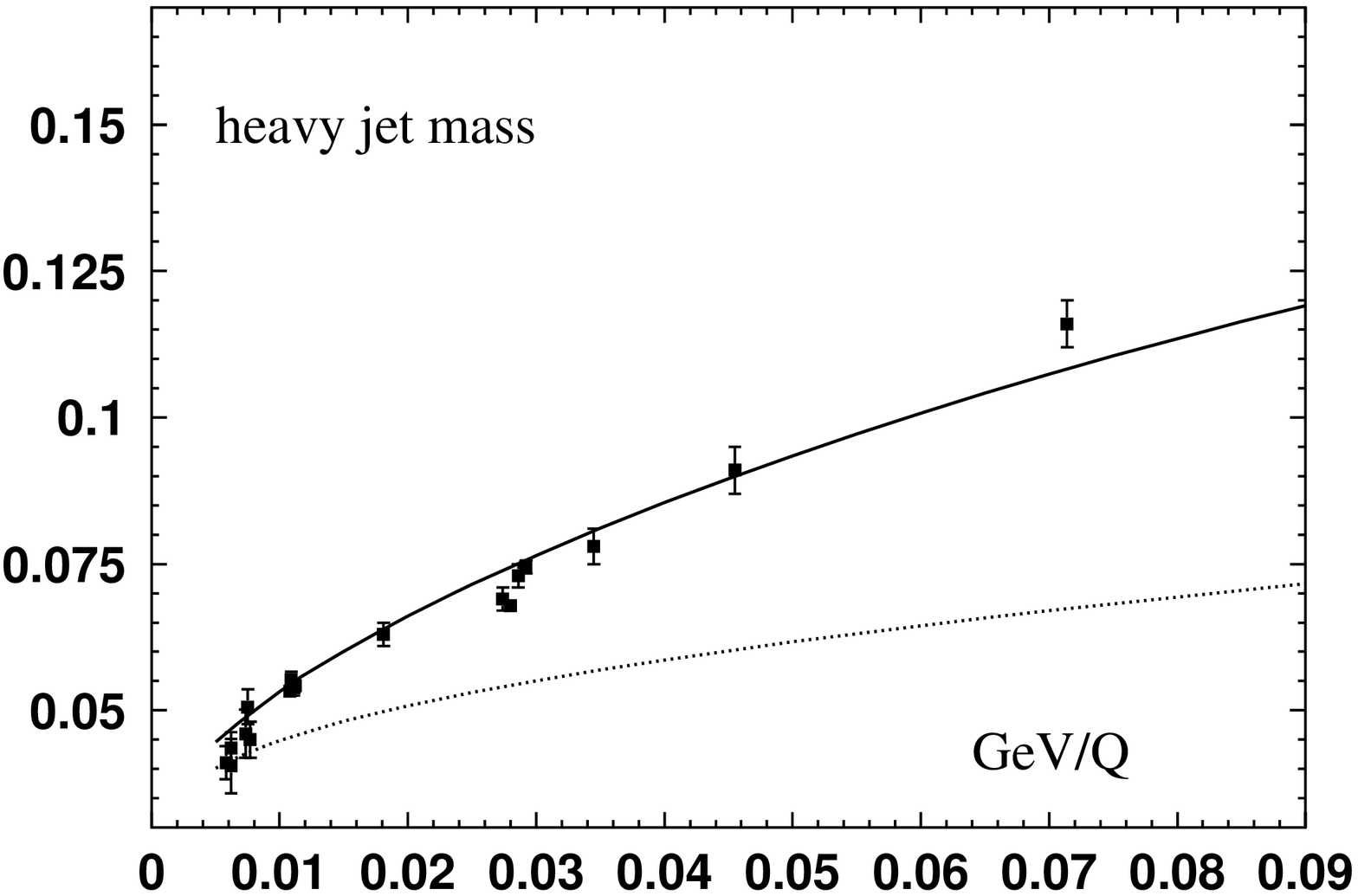}}
   \vspace*{-2.7cm}
\caption[dummy]{\small Energy dependence of 
$\langle 1-T\rangle$ (upper panel) and 
the heavy jet mass $\langle M_H^2/Q^2\rangle$ 
(lower panel) plotted as function 
of $1/Q$.\,\cite{Jad98} 
Dotted line: second-order perturbation 
theory with scale $\mu=Q$. Solid line: second-order perturbation theory 
with power correction added according to Eq.~(\ref{ksmu}) and with 
$\mu=Q$, $\mu_I=2\,$GeV. For $\bar{\alpha}_0(2\,\mbox{GeV})$ the fit values 
0.543 for thrust and 0.457 for the heavy jet mass\,\cite{Jad98} 
are taken. (Note that this reference uses $c_{1-T}=c_{M_H^2/s}=1$.) 
The dashed line shows second order perturbation theory at the very 
low scale $0.07 Q$ with no power correction added. For both 
observables $\alpha_s(M_Z)$ has been fixed to 0.12.  \label{fig20}}
\end{figure}

In Fig.~\ref{fig20} we compare the energy dependence of 
$\langle 1-T\rangle$ and the heavy jet mass 
$\langle M_H^2/Q^2\rangle$ with the prediction 
with and without 
a $1/Q$ power correction. It is clearly seen that (a) the second-order 
perturbative result with scale $\mu=Q$ is far too small and 
(b) the difference with the data points is fitted well by a
$1/Q$ power correction. 

In absolute terms the power correction added to thrust and the heavy jet 
mass is about $1\,\mbox{GeV}/Q$. This is a sizeable correction of order 
$20\%$ even at the scale $M_Z$, because the perturbative contribution is 
of order $\alpha_s(M_Z)/\pi$. The fit for $\bar{\alpha}_0$ is sensitive 
to the choice of renormalization scale $\mu$ and in general to the 
treatment of higher-order perturbative corrections. There is nothing 
wrong with this, because the very spirit of the renormalon approach is 
that perturbative corrections and non-perturbative hadronization 
corrections are to some extent inseparable. Hence we find it plausible 
that the $1/Q$ power correction accounts in part for large higher-order 
perturbative corrections, which are large precisely because they receive 
large contributions from IR regions of parton momenta. 
It has been noted\,\cite{BB96} 
that choosing a small scale, $\mu=0.13Q$, reduces the 
second order perturbative contribution and power correction 
significantly for $\langle 1-T\rangle$. In Fig.~\ref{fig20} (dashed 
curve) we have taken a very low scale, $\mu=0.07Q$, to illustrate the 
fact that the running of the coupling at this low scale can fake 
a $1/Q$ correction rather precisely (a straight line in the figure). 
Note that 
an analysis of $\langle 1-T\rangle$ in the 
effective-charge scheme\,\cite{CGM98} selects almost the same scale 
$\mu=0.08 Q$. A simultaneous fit\,\cite{CGM98} of 
$\alpha_s$, a third-order perturbative 
coefficient and a $1/Q$ power correction then leads to 
reduced power correction of order $(0.3\pm 0.1)\,\mbox{GeV}/Q$ 
consistent with the above argument.

\begin{figure}[t]
\vspace*{0.2cm}
   \epsfysize=7cm
   \epsfxsize=10cm
\vspace*{-0.2cm}
   \centerline{\epsffile{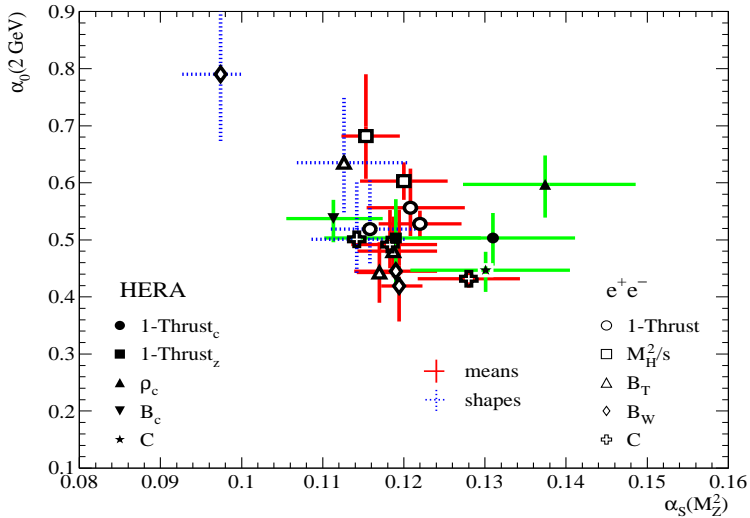}}
\caption[dummy]{\small 
Fit results of $\bar\alpha_0$ and $\alpha_S(M_{\mathrm{Z}})$.\,\cite{Bie00}
 \label{a0}}
\end{figure}

A comprehensive compilation\,\cite{Bie00} of the combined fits of the 
the parameter $\bar{\alpha}_0(\mu_I)$ and the strong coupling
$\alpha_S(M_{\mathrm{Z}})$ to the data on several event shapes
in $e^+e^-$-annihilation and deep inelastic scattering is 
shown in Fig.~\ref{a0}. The extracted values 
of $\bar{\alpha}_0$ center around   
$\bar{\alpha}_0(2$~GeV$) \sim 0.5$  that gives some support 
to the universality hypothesis. Its theoretical status has not been 
completely elucidated so far, different and somewhat conflicting 
arguments have been given.\,\cite{DW95,AZ95a,KS95b} 
In order to get further insight, 
a detailed analysis of IR sensitive contributions to the matrix 
elements for the emission of two partons\,\cite{DLMS97,DLMS98} has been 
undertaken. It was found that
for $\langle 1-T\rangle$, the jet masses, and the $C$-parameter the 
coefficient of the $1/Q$ power correction that is obtained for one gluon 
emission is rescaled by the {\em same} factor, 
often called the ``Milan factor'', whose value was later revised to
1.49.\,\cite{DMS99,Dok99}  
 The universality of the Milan factor for a class of event shapes 
is a direct consequence of the assumed 
dominance  of soft-gluon radiation
coupled with an underlying geometrical universality (linearity in 
transverse momenta of emitted gluons) in the shape variables 
themselves.\,\cite{SZ99}  
Note that the numerical value of the ``Milan factor'' is actually not 
important, since it is likely to be modified by yet higher-order 
corrections. The {\em universality of the ``Milan factor''} is hence 
the more important observation, since it allows us to relate 
different observables. Nevertheless, 
the analysis of soft-gluon effects at the 2-loop 
order is very interesting, since it is represents one of 
the few cases where power corrections have been explicitly investigated 
beyond one-gluon exchange for observables that do 
not have an operator product expansion.

\subsubsection{Event Shape Distributions}

The structure of power corrections to event shape distributions 
$(1/\sigma_{tot})d\sigma/dS$ is more complex. (The distributions 
are defined so that the mean value is given by 
$\langle S\rangle = (1/\sigma_{tot})\int_0^{S_{max}} \!\!dS\, 
S\, d\sigma/dS$.) In the following we consider event shapes defined 
so that $S\to 0$ corresponds to the two-jet limit. 
For small values of the event shape
variable $S$ the dynamics of soft-gluon emission depends on two 
different IR scales $QS$ and $\sqrt{Q^2S}$, and  
$1/Q \ll 1/Q\sqrt{S} \ll 1/(QS)$. The smallest scale of order $QS$
is related to the typical energy carried by soft gluons, while
the scale $Q\sqrt{S}$ defines the transverse momenta of the jets,
$k_T^2 \sim Q^2 S$. By examining the  
sensitivity of perturbative emission of soft gluons
with energy of order $QS$ and collinear particles with transverse 
momentum of order $Q^2 S$, we are lead to suspect non-perturbative 
corrections suppressed by powers of both scales. Then, since in the 
end-point region $S\sim \Lambda/Q$, we can expand the distributions
in powers of the larger scale $Q^2S$ and keep the leading term only
that corresponds to the resummation of all corrections 
of order $(1/QS)^k$, neglecting corrections 
of order $(1/Q\sqrt{S})^k$.
The resummation introduces the important concept of a shape function 
for soft-gluon emission,\,\cite{KS99,KT00} which we shall review
briefly on the  particular example of the thrust distribution 
$t\equiv 1-T$.\,\cite{KS99}

The starting point is the observation that the differential
thrust distribution $d\sigma/dt$ in the small $t$ region computed by 
resumming an infinite number of soft-gluon emissions is expected to
exponentiate under a Laplace transformation  
\begin{equation}
   \frac{1}{\sigma_{tot}} \int_0^{t_{\rm max}}\! dt \,e^{-\nu t}\,
   \frac{d\sigma}{dt} = e^{-S(\nu,Q)}
\label{thrust1}
\end{equation}      
with the exponent $S(\nu,Q)$ of the general form
\begin{eqnarray}
 S(\nu,Q) &=& \int_0^1\frac{dx}{x}
 \left(1-e^{-\nu x}\right)
\nonumber\\
&&\hspace*{-1.5cm}\times 
 \left(\int_{x^2Q^2}^{x Q^2}\!\! \frac{dk_\perp^2}{k_\perp^2}
 \Gamma[\alpha_s(k_\perp^2)] + B[\alpha_s(x Q^2)]
  +C[\alpha_s(x^2 Q^2)]\right).
\label{thrust2}
\end{eqnarray}
Here $\Gamma[\alpha_s]$ is the universal cusp anomalous dimension that 
controls soft {\em and} collinear gluon emission. 
(Compared to the original discussion\,\cite{KS99} 
we have added the additional function 
$C[\alpha_s(x^2 Q^2)]$ to the exponent. This function does not 
appear to the next-to-leading logarithmic accuracy, but we are not aware 
of a theoretical argument that would not allow terms of this structure 
in higher-orders. The following discussion proceeds under the
assumption that $C[\alpha_s(x^2 Q^2)]$ does not have a divergent 
perturbative expansion, as it can happen in  
Drell-Yan production.\,\cite{BB95b}) The next step  
separates the contribution of soft gluons introducing a cutoff in 
transverse momentum $k_\perp<\mu$. Heuristically, the contribution of 
gluons with  $k_\perp>\mu$ has to be defined as a perturbative contribution
$S_{PT}(\nu,Q,\mu)$ to $S(\nu,Q)$, and contributions of 
$k_\perp<\mu$ have to be 
promoted to the non-perturbative correction, $S_{NP}(\nu,Q,\mu)$,
so that $ S(\nu,Q) = S_{PT}(\nu,Q,\mu) + S_{NP}(\nu,Q,\mu)$. 
From the $\Gamma$-term in Eq.~(\ref{thrust2}), expanding formally 
in powers of $\nu/Q$ and neglecting all powers 
of $\nu/Q^2$, we obtain\,\cite{KS99}
\begin{equation}
 S_{NP}(\nu,Q,\mu) \sim \sum_{n>0}\frac{1}{nn!}\left(
   \frac{-\nu}{Q}\right)^n \int\limits_0^{\mu^2}\! d k_\perp^2\, 
   k_\perp^{n-2}\Gamma[\alpha_s(k_\perp^2)].
\label{thrust3}
\end{equation}
The integrals over the cusp anomalous dimension should eventually
be substituted by (dimensionful) non-perturbative parameters.
Since the variable $\nu$ is conjugate to $t$, Eq.~(\ref{thrust3})  
effectively organizes all power corrections in $1/(tQ)$. 

If $t\gg \Lambda/Q$, then keeping the first term only in the sum in 
Eq.~(\ref{thrust3}) is sufficient, $S_{NP}(\nu,Q,\mu) 
\sim {\lambda_1\nu}/{Q}$. Assuming exponentiation as in 
Eq.~(\ref{thrust1}),
the non-perturbative correction amounts  in this case to a  shift 
in the resummed perturbative thrust distribution\,\cite{KS95a,DW97}
\begin{equation}
  \frac{d\sigma}{dt}(t)  \longrightarrow \frac{d\sigma}{dt}(t-\lambda_1/Q)
  + O(1/(tQ)^2)\,,  
\end{equation}    
 with $\lambda_1$ the same non-perturbative parameter that parametrizes 
the $1/Q$ correction to the mean thrust, i.e.\ 
$\langle t\rangle_{1/Q} \sim \lambda_1/Q$.

If, on the other hand, $t \sim \Lambda/Q$, then all terms in the sum in 
Eq.~(\ref{thrust3}) have to be kept. The infinite set 
of non-perturbative parameters corresponding to the increasing powers 
of $\nu/Q$ defines a $Q$-independent function $f_t(\epsilon;\mu)$ through
\begin{equation}
  \int_0^\infty\!d\epsilon\, \exp(-\nu\epsilon/Q)\, f_t(\epsilon;\mu)
 \equiv \exp(-S_{NP}(Q/\nu,\mu)). 
\end{equation}   
The function $f_t(\epsilon;\mu)$ parametrizes the energy spectrum 
of non-perturbative soft-gluon emission for the thrust observable, 
similar to the shape 
function introduced for the description\,\cite{BSUV94,Neu94,KS94a} 
of the endpoint region in  
inclusive heavy meson decays such as $B\to\gamma X_s$.
With this definition, the differential thrust distribution corrected 
for the non-perturbative effects takes the form\,\cite{KS99}
\begin{equation}
  \frac{1}{\sigma_{tot}}\frac{d\sigma}{dt} = Q f_t(tQ) R_{PT}(0,\mu) + 
\int_0^{tQ}\!d\epsilon\, f_t(\epsilon;\mu)
 \frac{1}{\sigma_{tot}}\frac{d\sigma_{PT}}{dt}\left(t-\epsilon/Q\right),
\end{equation}
where $R(0,\mu)$  is the 
Sudakov factor taking into account the contribution of virtual 
soft gluons with the energy above the cutoff $\mu$ and the subscripts
``PT'' indicate that the corresponding quantity is calculated in 
perturbation theory. The shape function induces a smearing of perturbative
gluon radiation by non-perturbative corrections. 
Note that shape functions depend, in general, on the particular event shape
observable. In addition to thrust, also the heavy jet mass and 
the C-parameter distributions have been studied.\,\cite{KT00}    
 
In contrast to heavy quark decay where non-perturbative corrections 
extend the photon spectrum beyond the perturbative boundary of phase
space 
due to the energy distribution of the heavy quark in the meson, 
the non-perturbative corrections to the thrust distribution have an opposite 
effect. They shift the distribution inside the perturbative window
$0<t<t_{max}$ and describe the 
``evaporation'' of the energetic jets in the final 
state, loosing energy for the emission of soft gluons.
\begin{figure}[p]
   \epsfysize=7.4cm
   \epsfxsize=10.3cm
   \centerline{\epsffile{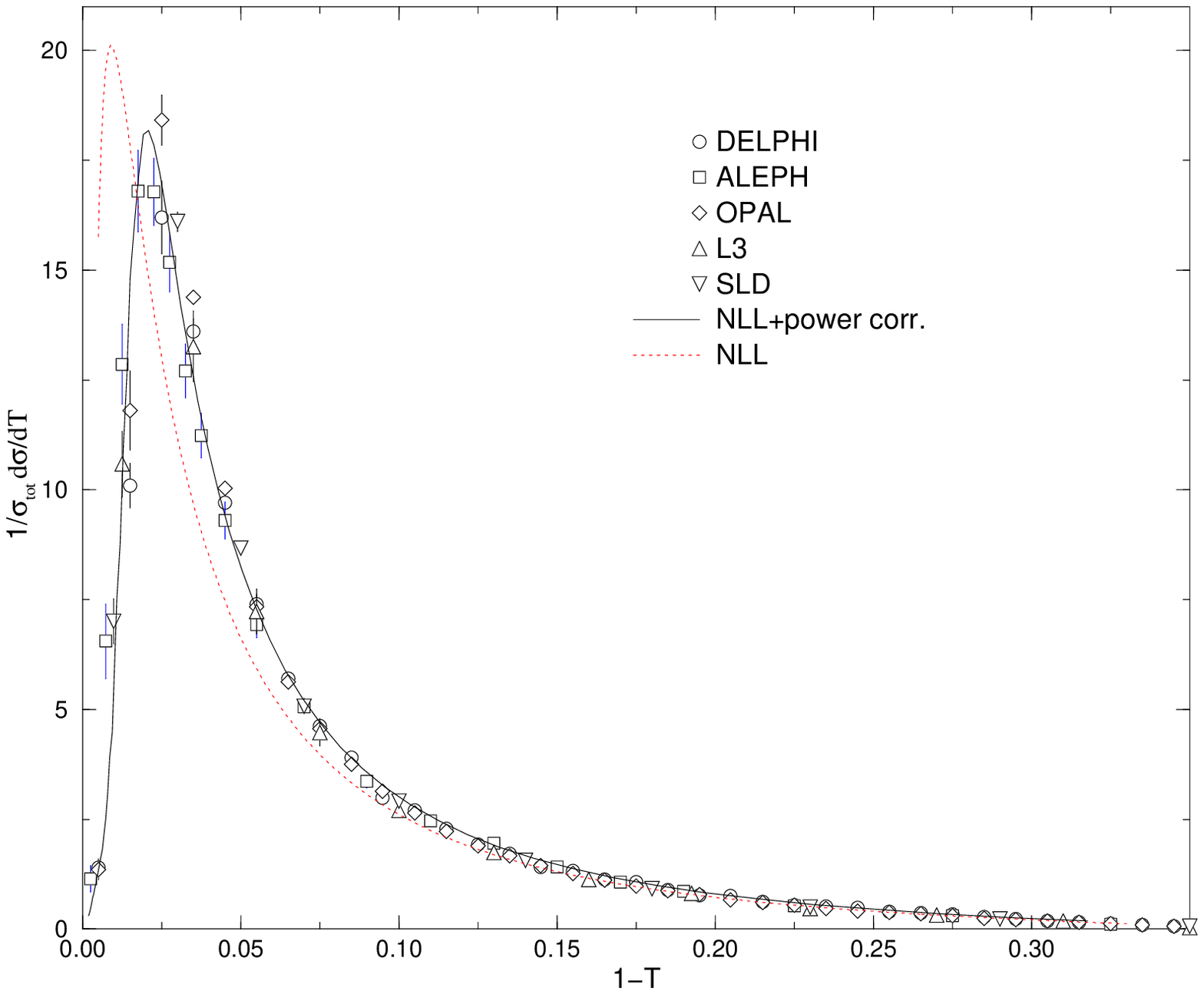}}
   \vspace{-1cm}
\caption[dummy]{\small 
 Fitting the shape function\,\cite{KS99} from the comparison with the data 
 on the differential thrust distribution at $Q=91.2$~GeV.
 \label{thrust-dif}}
   \vspace{1cm}
   \epsfysize=7.4cm
   \epsfxsize=10.3cm
   \centerline{\epsffile{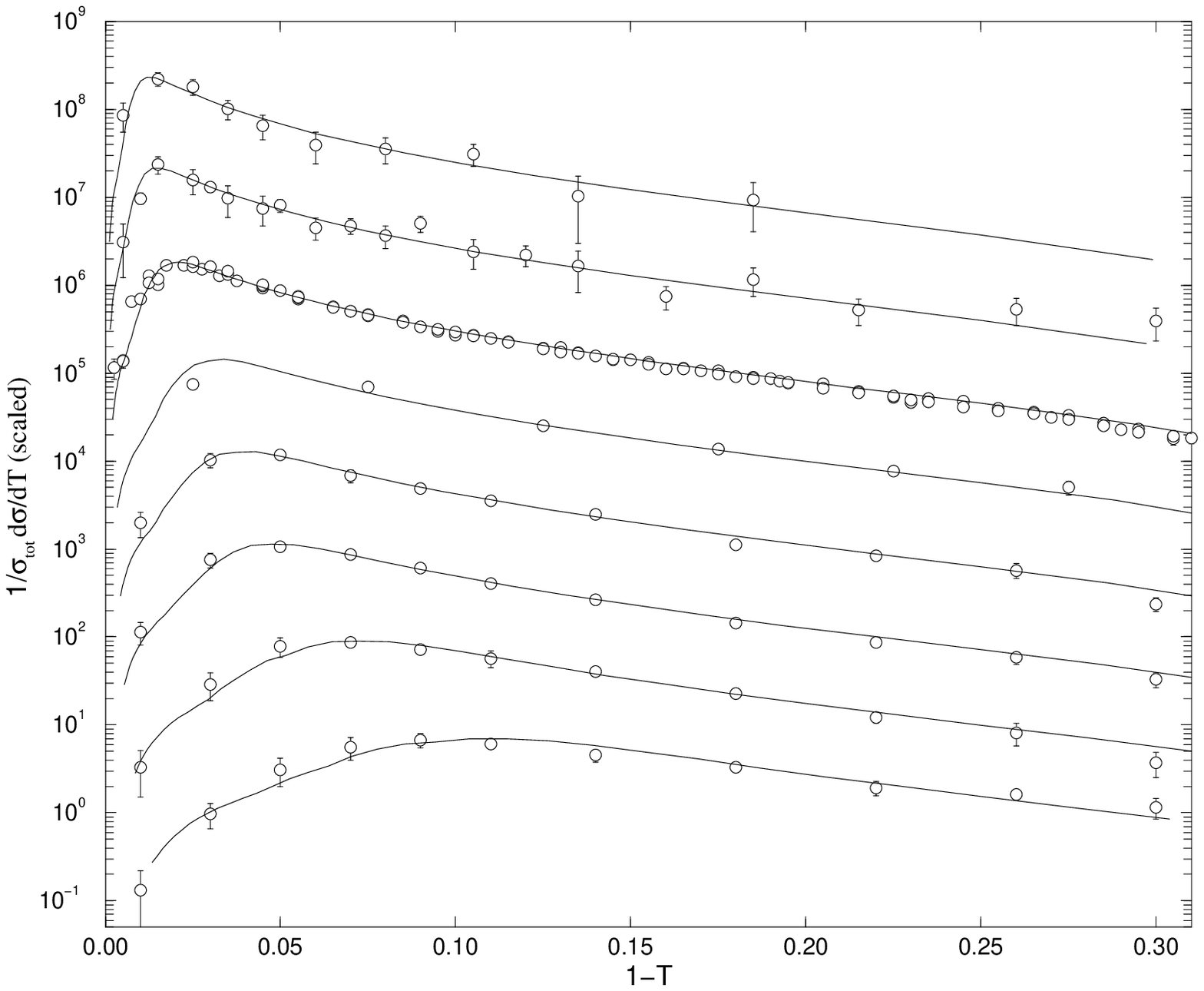}}
   \vspace{-1cm}
\caption[dummy]{\small 
 Using the shape function\,\cite{KS99} to predict 
 the differential thrust distribution at $Q=14,22,35,44,55,91,133,161$~GeV
 (from bottom to top). 
 \label{thrust-dif2}}
\end{figure}
In practice, one has to model the shape function using a certain ansatz
and fit the parameters to describe the thrust distribution at a certain 
value of the c.m. energy $s=Q^2$. The result of 
such a fit is shown in Fig.~\ref{thrust-dif}.
 The energy dependence of the
differential thrust distribution is then predicted without free 
parameters and appears to be in a good agreement with all available 
data, see Fig.~\ref{thrust-dif2}.

\subsubsection{The Energy Flow Correlation Function}

The most important statement that has emerged from the application 
of renormalons to event shapes and is supported by all existing 
evidence is that soft-gluon emission presents the 
only source of linear $\Lambda/Q$ power corrections. This 
observation has profound consequences since emission of soft gluons 
occurs at time scales that are much larger than those involved in 
the formation of perturbative narrow jets. As a consequence, these two 
subprocesses are quantum-mechanically incoherent and the soft-gluon 
distribution emitted by a pair of quark jets at wide angles depends
only on the direction and total color charge of the jets. 
This heuristic reasoning suggests factorization: to all orders in 
perturbation theory, inclusive cross sections for two narrow jets 
in $e^+e^-$ annihilation can be written as products of separate 
functions for the jets and the soft-gluon radiation, up to corrections
suppressed by powers of $(\Lambda/Q)^2$. As a consequence, we can use the 
eikonal approximation and replace the jets by eikonal, or Wilson lines
$W_{\pm} = {\rm Pexp}[ig_s\int_0^\infty du p_\pm^\mu A_\mu(up_\pm)]$ 
in which soft-gluon field is integrated along the light-like 
directions $p_{\pm}$ defined by the momenta of the outgoing jets. In this
approximation the shape function for a given event shape variable $S$ 
becomes 
\begin{equation}
   \frac{1}{\sigma_{tot}}\frac{d\sigma_{NP}}{dS} = 
 \sum_N |\langle N| W_+W_-^\dagger|0\rangle|^2 \delta(S-S(N)),
\end{equation}    
where the sum goes over all final states and an UV cutoff $\mu$ is 
assumed  for the gluon momenta.

The sum over the final states can be performed by introducing 
an operator ${\cal E}(\vec{n})$ that measures the density of energy flow 
in the direction of the unit vector $\vec{n}$ at spacial 
infinity.\,\cite{KS99,ST96}
Then, in particular, the shape function for the thrust distribution 
becomes\,\cite{KS99}
\begin{equation}
 f_t(\epsilon) = \langle 0|(W_+W_-^\dagger)(0)\,
   \delta\left(\epsilon - \int \! d\vec{n}\,w_t(\vec{n})\,{\cal E}(\vec{n},\mu)
\right)(W_- W_+^\dagger)(0)|0\rangle,
\end{equation} 
where $w_t(\vec{n})$ contains the information on the particular shape 
variable. In general, the 
complete information about soft-gluon emission is encoded 
in ``multi-energy flow'' correlation functions
\begin{equation}
  {\cal G}(\vec{n}_1,\ldots,\vec{n}_N;\mu) = 
  \langle 0|(W_+W_-^\dagger)(0)\,{\cal E}(\vec{n}_1),\ldots,{\cal E}(\vec{n}_N)
    (W_- W_+^\dagger)(0)|0\rangle
\end{equation}  
that describe the energy flow at spatial infinity.
Power corrections to different event shape averages can be calculated 
in terms of the energy flow as
\begin{equation}
 \langle S\rangle_{1/Q} = \int \! d\vec{n}\,
w_S(\vec{n}){\cal G}(\vec{n},\mu).
\end{equation}  
The concept of energy flow is very useful, since it allows us to
define the quantities of interest on an operator level, which makes
them more amenable to a rigorous analysis of power corrections. 
On the other hand, many important issues remain to be resolved such 
as the dependence of the energy flow correlation functions on 
the factorization scale $\mu$. But already at the present stage  
the concept of energy flow has provided insights into the extent to
which universality can be expected for power corrections to various 
moments or distributions of event shape observables.

\subsection{Heavy Quarks}
\label{heavyQ}

In this section we consider hard processes for which the large scale 
is given by the mass of a heavy quark. We discuss the notion 
of the (pole) mass of a heavy quark itself and its relation 
to the heavy quark potential, and consider
applications of these results to $t\bar t$ pair production near
the production threshold in $e^+ e^-$ annihilation.

\subsubsection{The Pole Mass}

Since quarks are not observed as asymptotic states, 
their masses generally have to be considered
as parameters in the QCD Lagrangian, on par with the strong coupling. 
Because the running of the coupling is conventionally considered using 
dimensional regularization, it is most natural to also employ the  
$\overline{\rm MS}$ scheme for the mass definition, introducing 
running quark masses $\bar{m}_q(\mu_{\overline{\rm MS}})$ and fixing their 
values at a certain reference scale. This is indeed the procedure 
used to deal with light quarks $q=u,d,s$ and also heavy quarks
provided the hard scale in the process is larger than or of order of the 
quark mass. On the other hand, using $\overline{\rm MS}$ heavy quark masses
is not convenient in processes where the hard scale is 
significantly smaller than the mass of the quark itself. 
The reason for this is that  the usual renormalization group 
expression 
\begin{equation}
   m_b(\mu) \simeq \left(\frac{\alpha_s(\mu)}{\alpha_s(m_b)}\right)^{4/b}
m_b(m_b)
\label{RGmass}
\end{equation}    
is physically irrelevant at $\mu\ll m_b$ because it is derived by assuming 
that $\mu$ is  the UV cutoff and thus the largest scale. 
(There is, {\em formally},  nothing wrong with taking $\mu\ll m_b$ 
in Eq.~(\ref{RGmass}). However, in calculations
of physical observables, e.g.\ heavy quark decay rates,  
we expect large perturbative corrections in higher 
orders in this case, because unphysical, large logarithms are
generated.)   
On the other hand, in the limit $\mu \ll m_b$ the heavy quark interacts
with gluons through the color Coulomb potential 
$V(\vec{r}) = - C_f\alpha_s(1/r)/|\vec{r}|$.
A restriction on the gluon momentum $|k| < \mu$ corresponds to the 
cutoff at large distances $r > 1/\mu $ so that the dependence 
of the mass parameter on 
$\mu$ at $\Lambda\ll \mu \ll m_b$ is in fact linear\,\cite{BU94,BSUV94,Ben98}
\begin{figure}[t]
   \epsfysize=4.8cm
   \epsfxsize=7cm
   \centerline{\epsffile{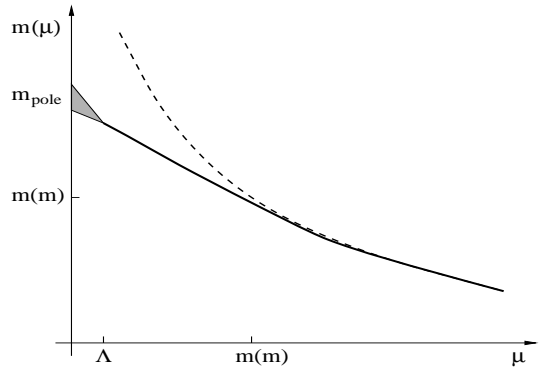}}
\caption[dummy]{\small 
A schematic representation of the scale dependence of a heavy quark 
mass defined with a ``physical'' IR cutoff (solid line) and in the 
dimensional regularization (dashed line). The shaded area illustrates
uncertainties of the mass definition when the scale is driven to  
values of order $\Lambda$  
 \label{mass}}
\end{figure}
\begin{equation}
   m_b(0)-m_b(\mu) \sim  C_f\alpha_s \mu\,, 
\end{equation}  
see Fig.~\ref{mass} for an illustration.  
The quantity $m(0)$ would correspond to a ``physical'' quark mass if it 
existed and in perturbation theory can be defined as the location 
of the pole in the perturbative quark propagator, i.e.\ the pole mass:
\begin{equation}
  m_{\rm pole} = m_{\overline{\rm MS}}(m_{\overline{\rm MS}})
   \left[1+\sum_n r_n \alpha_s^n(m_{\overline{\rm MS}})\right] 
\label{polemass}
\end{equation} 
The pole mass is IR finite, gauge independent and independent on the 
renormalization scheme.\,\cite{Tar81,Kro98} The perturbative 
series in Eq.~(\ref{polemass}) 
is, however, divergent:\,\cite{BB94a,BSUV94,Ben95}
\begin{equation}
    r_n \sim \mbox{\rm const}\cdot(2\beta_0)^n
\Gamma[n+1+\beta_1/(2\beta_0)^2].
\end{equation}
The sum of the series is ambiguous by an amount of order  $\Lambda$,
and, therefore, the quark pole mass is perturbatively defined 
only to an accuracy
\begin{equation}
   \delta m_{\rm pole} \sim \Lambda.
\end{equation} 
This uncertainty has important practical implications as it means that 
the pole mass has to be eliminated as a parameter in calculations of 
physical observables, if these 
observables are less sensitive to the IR region than the 
pole mass itself. 
Important examples of such observables are inclusive heavy quark 
decays\,\cite{BSUV94,BBZ94,NS95z,LMS95,BBB95b,BSU97} and 
top quark production near threshold 
in $e^+e^-$ annihilation. The second example will be discussed 
in some detail below. IR renormalons have also been investigated 
for exclusive heavy quark decays,\,\cite{NS95z,LMS95} in which a 
relation with the heavy quark pole mass emerges through the 
binding energy $\bar{\Lambda}$, defined in heavy quark effective theory.

\subsubsection{The Heavy Quark Mass with IR Subtractions.}

In order to deal with a well-defined non-perturbative parameter and
at the same time to retain the attractive features of the pole mass 
we have to perform an explicit scale separation. Heuristically, 
this would mean that the perturbative contributions in Eq.~(\ref{polemass})
have to be calculated using a cutoff $\mu$ at small momenta. 
The dependence of such subtracted mass on $\mu$ is linear and the 
coefficient depends on the particular procedure to implement the
IR-cutoff, but it is difficult in practice to implement such a 
procedure, and to maintain gauge invariance in particular. 

One suggestion\,\cite{Ben98} has been to utilize a close connection 
between ambiguities in the pole mass and the static 
heavy quark potential. 
The starting observation\,\cite{Ben98} is that the leading 
IR power correction 
to the potential in momentum space cannot be 
$\Lambda/|\vec{q}\,|$, but has to be quadratic:
\begin{equation}
\label{potmom1}
\tilde{V}(\vec{q}\,) = -\frac{4\pi C_F\alpha_s(\vec{q}\,)}{\vec{q}^{\,2}}
\left(1+\ldots + \,{\rm const}\cdot\frac{\Lambda^2}{
\vec{q}^{\,2}} + \ldots\right).
\end{equation}
To avoid misunderstanding, note that we are not concerned with the 
long-distance behaviour of the potential at $q\sim \Lambda$, 
but with the leading power corrections of the form $(\Lambda/q)^k$,  
which correct the perturbative Coulomb potential when $q$ is still 
large compared to $\Lambda$.

When we consider the coordinate space potential, given by the Fourier 
transform of $\tilde{V}(\vec{q}\,)$, a new situation arises.
It is easy to see by dimensional analysis that the contribution of small 
$|q|<\mu $ in the Fourier integral is a $r$-independent constant:
\begin{equation}
   \int\limits_{|\vec{q}\,|<\mu_f} 
\!\!\!\frac{d^3\vec{q}}{(2\pi)^3}\,e^{iq\cdot r}\tilde{V}(\vec{q}\,)
 \sim 
\int\limits_{|\vec{q}\,|<\mu_f} 
\!\!\!\frac{d^3\vec{q}}{(2\pi)^3}\,\tilde{V}(\vec{q}\,) = 
  \mbox{\rm const} \cdot \mu\,.   
\end{equation}  
This implies a long-distance and hence non-perturbative 
correction
\begin{equation}
V(r) = -\frac{C_F\alpha_s(1/r)}{r}\left(
1+\ldots + \,{\rm const}\cdot\Lambda r + 
\ldots\right)
\end{equation}
for the coordinate space potential that can also be observed
through the calculation of renormalons generated by one-gluon 
exchange with vacuum polarization insertions.\,\cite{AL95,AZ97b} 

The $r$-independent constant in the potential is closely 
related to the uncertainty in the pole mass. This is easy to understand
since  the total energy of a heavy quark-antiquark system 
$E(r)= 2 m_{\rm pole}+V(r)$ is a physical observable and has to be well
defined in the heavy quark limit.\,\cite{Ben98,HSSW98} We can then
define a potential-subtracted (PS) quark mass and a 
subtracted potential as\,\cite{Ben98}
\begin{eqnarray}
  m_{\rm PS}(\mu_f) &=& m_{\rm pole}-\delta m(\mu_f).
\nonumber\\
  V(\vec{r},\mu_f) &=& V(\vec{r}\,)+2\delta m(\mu_f)
\end{eqnarray} 
where 
\begin{equation}
\label{deltam}
\delta m(\mu_f) = -\frac{1}{2}\int\limits_{|\vec{q}\,|<\mu_f} 
\!\!\!\frac{d^3\vec{q}}{(2\pi)^3}\,\tilde{V}(\vec{q}\,).
\end{equation}
To 2-loop accuracy, the relation 
of the PS mass to $\bar{m}\equiv 
\bar{m}_{\overline{\rm MS}}( \bar{m}_{\overline{\rm MS}})$ 
is given by 
\begin{eqnarray}
\label{pstoms}
m_{\rm PS}(\mu_f) &=& \bar{m}\,\Bigg\{1+\frac{4\alpha_s(\bar{m})}{3\pi} 
\left[1-\frac{\mu}{\bar{m}}\right] 
\nonumber\\
&&\hspace*{-1.2cm}+ \left(\frac{\alpha_s(\bar{m})}{\pi}
\right)^2\left[K_1-\frac{\mu}{3 \bar{m}}\left(K_2-4\pi\beta_0\left[
\ln\frac{\mu^2}{\bar{m}^2}-2\right]\right)\right]+\ldots\Bigg\},
\end{eqnarray}
where $K_1=13.44-1.04 N_f$ is the 2-loop coefficient in the relation 
of $m_{\rm pole}$ to $\bar{m}$\,\cite{GBGS90}  
and $K_2=10.33-1.11 N_f$ the 1-loop correction to the Coulomb 
potential in momentum space. (The 3-loop relation is also known.) 
The important point is that in the perturbative expansion that 
relates the two masses in Eq.~(\ref{pstoms}) the leading 
IR renormalon divergence has been eliminated. In practice, this 
leads to smaller perturbative coefficients starting already at 
two loops.

We can use the PS mass and subtracted potential instead of the 
pole mass and the Coulomb potential to perform Coulomb resummations 
for threshold problems. The benefit of using an unconventional 
mass definition is that large perturbative corrections related to 
strong renormalon divergence associated with the coordinate 
space potential are obviated. Physically, the crucial point is 
that, contrary to intuition, heavy quark cross sections near threshold 
are in fact less long-distance sensitive than the pole mass 
and the coordinate space potential. The cancelation is made 
explicit by using a less long-distance sensitive mass definition.  

\subsubsection{Top Quark Production in $e^+e^-$ Annihilation} 

It is perhaps surprising that the renormalon divergence in the pole 
mass/$\overline{\rm MS}$ mass relation and the mass subtractions 
discussed above have played a very important role for the top quark, 
which is so much heavier than the QCD scale. The reason for this is 
the extraordinary precision of about $100\,$MeV 
with which the top quark mass can in principle be 
determined by scanning the pair production threshold at an 
$e^+ e^-$ collider. 

The improvement in convergence due to the subtraction term can be 
seen on the one hand in the perturbative conversion to the 
$\overline{\rm MS}$ mass, and in the $t\bar{t}$ line shape on 
the other hand.  Numerically, 
the series that relate the pole and PS mass, respectively, 
to the $\overline{\rm MS}$ mass $\bar{m}_t=\bar{m}_t(\bar{m}_t)$ 
are as follows:
\begin{eqnarray}
&&\hspace*{-0.65cm}
m_t = \big[165.0+7.64+1.64+0.52
+ 0.25\,(\mbox{est.})\big]\,
\mbox{GeV}
\label{polerel}\\
&&\hspace*{-0.65cm}
m_{t,\rm PS}(20\,\mbox{GeV}) = \big[165.0+6.72+1.21+
0.29+0.08\,(\mbox{est.})\big]\,
\mbox{GeV}.
\label{psrel}
\end{eqnarray}
for $\bar{m}_t=165\,$GeV and $\alpha_s(\bar{m}_t)=0.1091$
(corresponding to $\alpha_s(m_Z)=0.119$). 
The 3-loop coefficients can also be computed using recent 
results.\,\cite{MvR00,CS99} 
The 4-loop estimate uses the ``large-$\beta_0$'' limit.\,\cite{BBB95} 
The improved convergence is evident and significant on the scale 
of $100\,$MeV set by the projected uncertainty on the 
mass measurement.

\begin{figure}[p]
   \vspace{-2.5cm}
   \epsfysize=18cm
   \epsfxsize=13cm
   \centerline{\epsffile{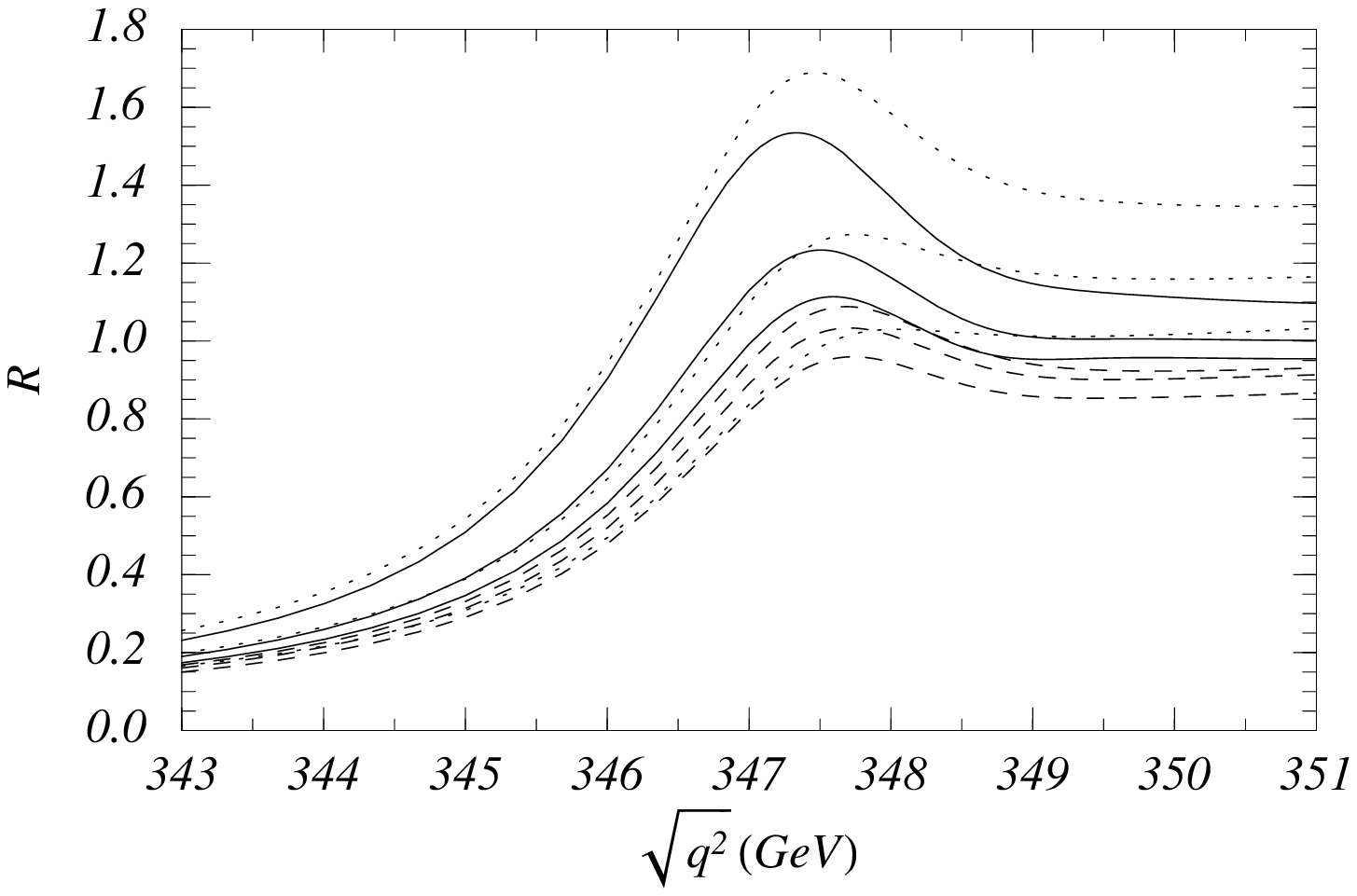}}
   \vspace*{-11.7cm}
   \epsfysize=18cm
   \epsfxsize=13cm
   \centerline{\epsffile{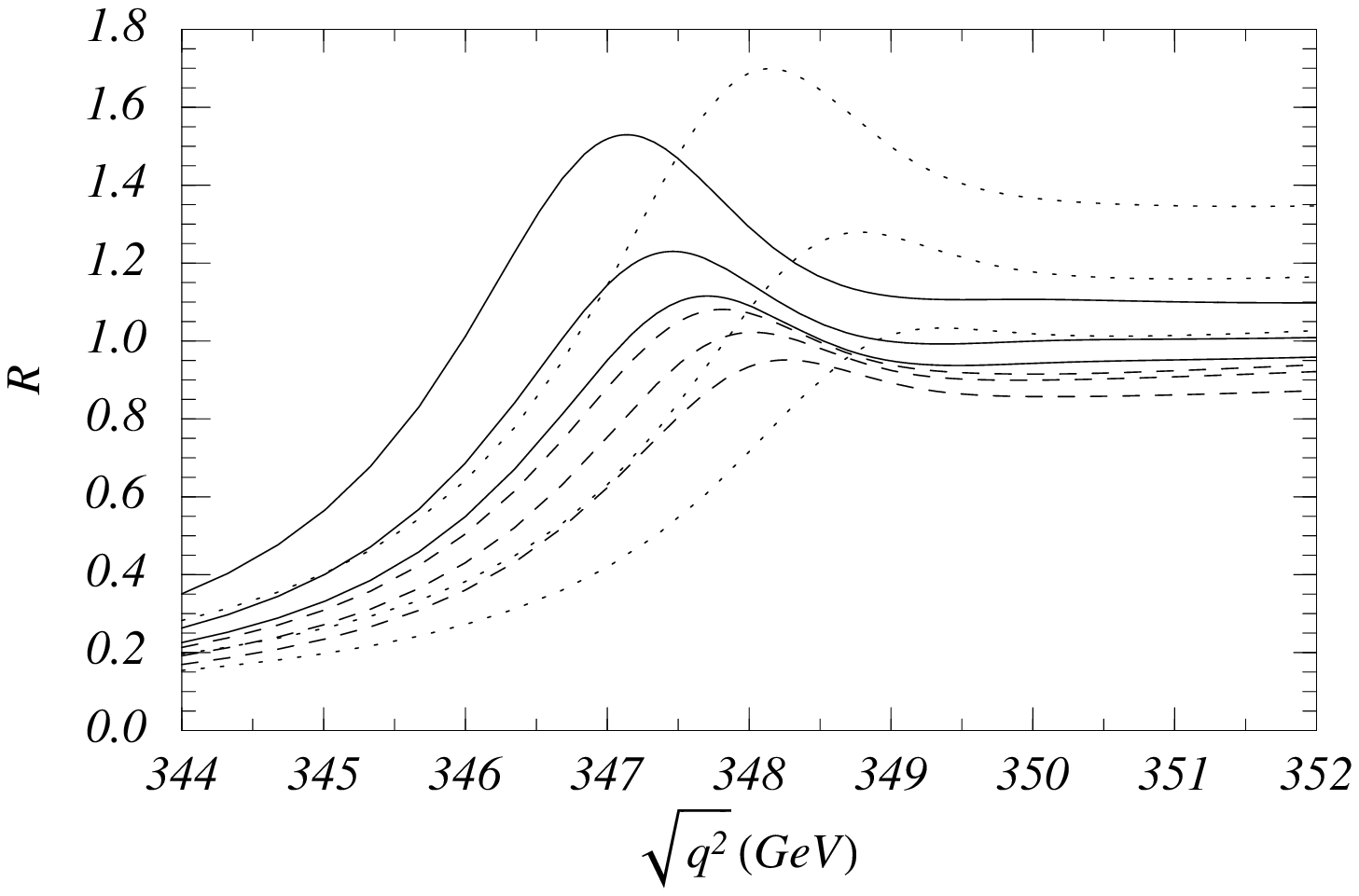}}
   \vspace*{-10cm}
\caption[dummy]{\small 
The total normalized $t\bar{t}$ cross section (virtual 
photon contribution only) in the threshold region at 
leading order (dotted), next-to-leading order (dashed) and 
next-to-next-to-leading order (NNLO) versus the center-of-mass 
energy.\,\cite{BSS99} Both plots use 
$\bar{m}_t(\bar{m_t})=165\,$GeV, $\Gamma_t=1.43\,$GeV 
and $\alpha_s(m_Z)=0.119$ as input. The three curves for each case 
refer to the renormalization scale 
$\mu=\left\{15 (\mbox{upper}); 30 (\mbox{central}); 
60 (\mbox{lower})\right\}\,$GeV. The PS mass corresponding 
to $\bar{m}_t(\bar{m_t})=165\,$GeV is
$m_{t,\rm PS}(20\,\mbox{GeV})=173.30\,$GeV, the pole mass 
$m_{t,{\rm pole}}=175.05\,$GeV.
The upper panel shows the 
successive approximations in the PS mass scheme, 
the lower panel in the pole mass scheme. 
\label{figtop}}
\end{figure}

The calculation of the top quark pair production cross section near 
threshold uses a non-relativistic effective theory to accomplish 
the all-order resummation of Coulomb ``singularities''. We cannot 
go into details here except to mention that the leading 
order cross section\,\cite{FK87} can be obtained from the 
effective Lagrangian 
\begin{eqnarray}
\label{pnrqcd}
{\cal L}_{\rm eff}
 &=& 
\psi^\dagger \left(i \partial^0+\delta m_t(\mu_f)+
\frac{\vec{\partial}^2}{2 m_t}+i\Gamma_t
\right)\psi 
\nonumber\\
&&\hspace*{0cm}
+ \,\chi^\dagger \left(i \partial^0+\delta m_t(\mu_f)
-\frac{\vec{\partial}^2}{2 m_t}+i\Gamma_t
\right)\chi
\nonumber\\
&&\hspace*{0cm}
+\, \int d^{d-1} r \left[\psi^\dagger\psi\right](r)
\left(-\frac{C_F\alpha_s}{r}\right) 
\left[\chi^\dagger\chi\right](0),
\end{eqnarray}
where $\psi$ and $\chi$ denote non-relativistic quark and antiquark 
fields, respectively. The leading order Coulomb potential is part of 
the leading order Lagrangian and cannot be treated as a 
perturbative interaction term. The effective Lagrangian yields 
the Schr\"odinger equation for the $t\bar{t}$ system, the only 
peculiarity being that the Schr\"odinger equation contains a 
non-hermitian term to account for top quark decay and a residual mass 
term $\delta m_t(\mu_f)$ defined by Eq.~(\ref{deltam}). The residual 
mass implies that the non-relativistic static energy is $\sqrt{q^2}
-2 m_{t,\rm PS}(\mu_f)$ rather than $\sqrt{q^2}
-2 m_t$ and by expressing the cross section in terms of 
$\sqrt{q^2}-2 m_{t,\rm PS}(\mu_f)$, the convergence of the perturbative 
approximation is also improved here. 

The cross section near threshold is now known to 
next-to-next-to-leading order\,\cite{HT98,KY98,Yak99,BSS99,NOS99} 
and it is at this 
order that the improvement of convergence of the peak of the 
line shape becomes particularly visible,\,\cite{BSS99,HT99,All} as 
shown in Fig.~\ref{figtop}. Subtracted top quark 
masses are therefore essential to take profit from the small 
experimental error. As shown above the subtracted masses can 
be accurately converted to the $\overline{\rm MS}$ mass which 
should then be a useful standard. Subtractions 
different\,\cite{BSUV97,HLM99} 
from the potential subtraction can be conceived and 
are useful as far 
as they eliminate the leading infrared sensitive term from 
the pole mass. They have also been used in top quark 
production.\,\cite{HT99,All}

\subsection{Synopsis of Other Results}

The analysis of renormalon divergences can be used to study 
non-perturbative corrections
to many other hard QCD processes. The following gives an incomplete 
discussion of some of these processes. 
\vskip0.2cm

{\em Fragmentation in $e^+e^-$ annihilation.}~Inclusive 
single particle production in $e^+e^-$ annihilation is 
the time-like
analogue of deep inelastic scattering. A renormalon 
analysis\,\cite{DW97,DW97a,BBM97}
predicts the leading power corrections to the differential cross
section to be of order $(\Lambda/Q)^2$,
in agreement with the light-cone expansion,\,\cite{BB91} 
and suggests the following parametrization:
\begin{eqnarray}
\frac{d\sigma_L^{\rm tw-4}}{dx}(x,Q^2) &=& 
\frac{\Lambda^2}{Q^2}\!\int_x^1\!\frac{dz}{z}\,
\Bigg\{c_{q}^{L}\left[\delta(1-z)+\frac{2}{z}
\right]D_q(x/z,\mu) \nonumber\\
&&\,+ c_{g}^{L}\,\frac{1-z}{z^3}D_g(x/z,\mu)
\Bigg\}~,
\label{parL}
\\
\frac{d\sigma_{L+T}^{\rm tw-4}}{dx}(x,Q^2) &=& 
\frac{\Lambda^2}{Q^2}\!\int_x^1\!\frac{dz}{z}
\Bigg\{c_{q}^{L+T}\!\left[\frac{-2}{[1-z]_+} + 1 + 
\frac{1}{2}\delta'(1-z)\right]D_q(x/z,\mu) 
\nonumber\\
&&\,+ \left[c_{g}^{L+T}\,\frac{1-z}{z^3} + d\right]D_g(x/z,\mu)
\Bigg\},
\label{parLT}
\end{eqnarray} 
where $D_{i}$ denotes the leading-twist fragmentation function for parton 
$i$ to decay into any hadron, ``$L$+$T$'' the sum of longitudinal 
and transverse fragmentation cross sections and the plus distribution 
is defined as usual. The power corrections are added to the 
leading-twist cross sections as 
\begin{equation}
\frac{d\sigma_P}{dx}(x,Q^2) = \frac{d\sigma_P^{\rm tw-2}}{dx}(x,Q^2) + 
\frac{d\sigma_P^{\rm tw-4}}{dx}(x,Q^2).
\end{equation}
The constants $c_k$ and $d$ are to be fitted to data and depend 
on the order of perturbation theory and factorization scale $\mu$ 
adopted for the leading-twist prediction. 

Owing to energy conservation, the parton fragmentation functions 
disappear from the second moments
\begin{equation}
\label{moment2}
\sigma_{L,T} \equiv  \sum_H\frac{1}{2}\int_0^1\! 
dx \, x\frac{d\sigma_{L,T}^H}{dx},
\end{equation}
which can therefore be calculated in perturbation theory up to 
power corrections. 
The power expansion of the fragmentation cross section 
has strong soft-gluon singularities and the expansion parameter 
relevant at small $x$ is $\Lambda^2/(Q^2 x^2)$. This can be related to 
the fact that in perturbation theory the hard scale relevant to 
gluon fragmentation is not $Q$, but the energy $Q x$ of the fragmenting 
gluon. It was noted\,\cite{DW97a,BBM97} that these strong singularities 
lead to a linear $\Lambda/Q$ non-perturbative correction to the 
$\sigma_L$ and $\sigma_T$.
This can be seen from 
\begin{equation}
\int\limits_{\Lambda/Q} dx\,\frac{1}{2}\,x\left[
\frac{\Lambda^2}{Q^2 x^2}
\right]^n \sim \frac{\Lambda}{Q}
\end{equation}
for any $n$,  
which also tells us that the correct $1/Q$ power correction is 
obtained only after resumming the power expansion at definite $x$ 
to all orders. To the 2-loop accuracy, the IR contribution to 
$\sigma_L$ appears to be related to that for the $C$-parameter.\,\cite{DMS99}

The total cross section in $e^+ e^-$ annihilation 
into hadrons is given by the sum of the transverse and longitudinal 
cross section. In this sum all power corrections of order 
$1/Q$, $1/Q^2$ and $1/Q^3$ cancel as expected from the operator 
product expansion.
\vskip0.2cm

{\em Drell-Yan production.}~Drell-Yan production of a lepton pair or 
a massive vector boson, $A+B\to \{\gamma^*,W,Z\}(Q) + X$, where $X$ 
is any hadronic final state, presents the best studied case of a hard 
process with two disparate hard scales for which logarithmically 
enhanced contributions due to soft-gluon emission 
can be resummed systematically to all orders of perturbation 
theory.\,\cite{Ste87,CT89} In Mellin space, the cross section 
$d\sigma_{\rm DY} /dQ^2$ factorises into a product of parton distribution
functions and the quark-antiquark scattering cross section that
takes the form 
\begin{equation}
\label{omega} 
\omega_{q\bar{q}}(N,\alpha_s(Q)) = H(\alpha_s(Q))\, \exp\left[
E(N,\alpha_s(Q))\right] + R(N,\alpha_s(Q)),
\end{equation}
where $R(N,\alpha_s(Q))$ vanishes as $N\to \infty$, 
$H(\alpha_s(Q))$ is independent of $N$, and the 
exponent is given by 
\begin{eqnarray} 
\label{exponent} 
E(N,\alpha_s(Q)) \!\!&=&\!\!\int_0^1 d z\,\frac{z^{N-1}-1}{1-z}\,
\Bigg\{2\int_{Q^2 (1-z)}^{
Q^2 (1-z)^2}\!\! \frac{d k_t^2}{k_t^2}\,\Gamma[\alpha_s(k^2_t)]
\nonumber\\[-0.2cm]
&&\hspace*{+1cm}
{}+\,
B[\alpha_s((1\!-\!z) Q^2)]
+C[\alpha_s((1\!-\!z)^2 Q^2)]\Bigg\}.
\end{eqnarray} 
A similar expression is valid for the thrust distribution,
cf.\ Eq.~(\ref{thrust2}), in which case we argued that it implies 
a linear in $1/Q$ non-perturbative correction because of the   
corresponding IR sensitivity of the first term involving the integral 
of the QCD coupling over low scales. 
For Drell-Yan production it happens, however,
that the infrared renormalon contributions arising in this way\,\cite{CS94} 
are exactly canceled by the divergent perturbative expansion 
of the function $C[\alpha_s]$.\,\cite{BB95b}
 The cancelation was shown explicitly in the large-$\beta_0$ approximation
and can be interpreted\,\cite{BB95b} as the cancelation of $O(1/Q)$
corrections between soft-gluon emission at different angles.
In particular, the large-angle, non-collinear emission is important. 
This conclusion is general and extends beyond Drell-Yan process.
A recent reanalysis\,\cite{SV99} 
of soft-gluon resummation in Drell-Yan production
also arrives at the cancelation of the 
leading $1/Q$ IR contributions. 
The absense of $1/Q$ power corrections has been 
put\,\cite{AZ96a,AZ96b,ASZ97} into the more general 
context of Kinoshita-Lee-Nauenberg (KLN) cancelations. Knowing 
that any potential $1/Q$ correction would come from soft particles, 
but not collinear particles, the KLN transition amplitude can be constructed,  
which includes a sum over soft initial and final 
particles degenerate with the annihilating $q\bar{q}$ pair. 
The KLN transition amplitudes have no $1/k_0$ (where $k_0$ stands 
for the energies of the soft particles) contributions (collinear 
factorization is implicitly assumed). As a consequence, the amplitude 
squared, integrated unweighted over all phase space, is proportional to 
$dk_0 k_0$, which by power counting implies at most $1/Q^2$ 
power corrections. To make connection with a physical process, 
one has to demonstrate that the sum over degenerate initial states 
can actually be dispensed of. This can be shown in an abelian theory 
using Low's theorem. The generalization to QCD is still 
an open problem.
\vskip0.2cm

{\em Hard exclusive reactions.}~The theory of hard exclusive scattering 
is much less developed compared to inclusive processes. Also from 
the experimental side there is conflicting evidence. In this situation
simple estimates using renormalons can provide important insight.
In a generic hard exclusive process
one finds two sources of renormalon 
divergence and power corrections. The first is power corrections in 
the hard coefficient function, which are present independently of the 
form of the hadron wave function. These correspond to higher-twist 
corrections in the hard scattering formalism. Additional power 
corrections are generated after integrating with the hadron wave 
function over the parton momentum fractions and these 
depend on the details 
of the wave function. These power corrections 
arise from the region of small parton momentum fraction and can be 
associated with power corrections due to the ``soft'' or ``Feynman'' 
mechanism for exclusive scattering. For the simplest reaction\,\cite{GK97}
 $\gamma^*+\gamma \to \pi^0$
and the deeply virtual Compton scattering\,\cite{BS98}
 $\gamma^*+A\to \gamma+B$
both power corrections are of order $1/Q^2$. 
Another interesting application\,\cite{And00} concerns the structure of the 
light-cone expansion of a non-local operator sandwiched between vacuum
and the pion that is used to define the pion distribution amplitude
$\phi_\pi(u)$:
\begin{eqnarray}
 \langle 0| \bar d(0)\gamma_\mu\gamma_5 u(x)|\pi^+(p)\rangle \!&=&\!
  \!i f_\pi p_\mu \!\int_0^1\! du\,e^{-iupx}
   \bigg[
\bigg(1+\!\sum_n r_n(u)\alpha_s^n\bigg)\phi_\pi(u)
\nonumber\\&&
\hspace*{2cm} 
+ x^2 g_1(u)+O(x^4)\bigg]. 
\end{eqnarray} 
The function $g_1(u)$ is interpreted as a two-particle 
pion distribution function of twist-4\,\cite{BF90} and is usually 
estimated to be $g_1(u)\sim \mbox{\rm const}\cdot u^2(1-u)^2$ by  the  
contribution of the lowest conformal operator. The renormalon estimates
can give constraints on possible contributions of high orders in the 
conformal expansion and in the large-$\beta_0$ limit the result 
is\,\cite{And00} $g_1^{\rm ren}(u)\sim \mbox{\rm const}\cdot u(1-u)$.
This result is significant since the behavior of $g_1(u)$
at the end points determines the parametric size of power corrections
to inclusive reactions involving pions. In particular, a nonzero value 
$g_1(u\to 1)$ would invalidate factorization for the pion form factor.
The further softening of the endpoint behavior $(1-u)\to (1-u)^2$ 
predicted by the conformal expansion must be interpreted as the effect
of the hierarchy of the anomalous dimensions of conformal operators.   
\vskip0.2cm

{\em Deep inelastic scattering in the small-$x$ limit.}~ 
Exploratory studies\,\cite{Lev95,ARS96} 
have been performed in the context of small-$x$ structure functions  to 
use renormalons in order to obtain  a certain prescription to 
implement the running coupling in the BFKL equation. An apparent 
$1/Q$ correction to the kernel is suppressed\,\cite{ARS96}  after convolution 
with the hadron wave function such that the correction to 
the structure function is only of order $1/Q^2$.
More recently,  similar studies have been carried out involving the
next-to-leading order BFKL kernel.\,\cite{FL98,CC98} 
The series expansion of the solution to the BFKL equation with the exact 
1-loop running coupling produces a series expansion of the form\,\cite{KM98}
\begin{equation}
\sum_n\left(a y \alpha_s^3\right)^{n/2}\,\Gamma(n/2),
\end{equation}
where $a=42\zeta(3)\beta_0^2/\pi$ and $y$ is the (large) 
rapidity that characterizes a scattering process in the BFKL limit.
If we take the Borel transform with respect to $\alpha_s^3$, the
above series leads to a typical renormalon pole. The unusual 
feature is that location of the 
renormalon pole depends on the kinematic variable $y$, and  
not only in overall prefactor. When $a y \alpha_s^3\sim 1$ the series 
diverges from the outset and no perturbative approximation is possible. 
This leads to the constraint $y<1/(a \alpha_s^3)$ for 
rapidities to which the BFKL treatment can be applied, in
agreement with other methods.\,\cite{Mue97}


\section{Ultraviolet Renormalons}
\label{sec:uvrenormalon}

Perturbative expansions in renormalizable field theories exhibit
another kind of divergence known as the ultraviolet (UV) renormalon. 
This divergence is easy to identify in simple sets of diagrams. For 
example, the integral in Eq.~(\ref{eq:smallk}) has a contribution 
of order $(-\beta_0/q)^n n!$ from large $k\gg Q$, where $q$ depends 
on the behavior of $I(k)$ at large $k$. For UV finite or renormalized
quantities the leading term $I(k)\sim 1/k^4$ is absent or subtracted 
and $I(k) \sim 1/k^6$ in general. This leads to $q=1$. Higher orders 
in the large-$k$ expansion of $I(k)$ give rise to divergent behavior 
with larger $q$ and the corresponding singularities in the Borel 
transform. 

We note that since $k\gg Q$ the powers of the logarithm in 
Eq.~(\ref{eq:smallk}) alternate in sign. Hence the terms of the 
divergent series are also sign-alternating. For such series the 
Borel integral can be unambiguously defined and so it seems that 
we need not concern ourselves further with this type of divergence. 
However, it is not obvious that Borel summation is the correct 
way to recover the non-perturbative result. The series alone tells 
us only that we can approximate the exact result up to an uncertainty 
of order $(\Lambda/Q)^2$, if $q=1$ and if the series expansion is 
expressed in terms of $\alpha_s(Q)$.

This is a puzzling result. For the GLS sum rule $I_{\rm GLS}$, 
used as an example in the introduction, this implies an ambiguity 
from arbitrarily large loop momenta as large as the leading 
higher-twist correction, even though the strong coupling is small 
in the ultraviolet. And how could such an UV effect be non-local 
(non-polynomial) in the external momentum $Q$? We shall see later 
in this section 
that the estimate $(\Lambda/Q)^2$ is too naive and that there is
indeed no ambiguity associated with UV renormalons. But we shall first
discuss the theory of UV renormalons independent of particular sets 
of diagrams.

UV renormalons are related to internal loop momenta larger than any 
external momentum; they disappear if Green functions are computed with 
a UV cutoff not taken to infinity. This suggests that the methods 
of renormalization theory can be used to characterize UV renormalons. 
The simple example of the function $I(k)$ discussed above shows that 
the first UV renormalon ($q=1$) is related to dimension-6 operators,
since this is what the first non-vanishing term in the large-$k$
expansion would yield. 
The first UV renormalon is particularly simple since for dimensional 
reasons we can allow only a single insertion of a dimension-6 
operator. Parisi\,\cite{P78} has used this insight to compute the nature
of the leading UV renormalon singularity in the scalar
$\phi^4$-theory; the connection with effective theories at finite UV
cutoff was also pursued.\,\cite{BD84} Later particular sets of 
diagrams were investigated\,\cite{dCP95,VZ94,BS96,PdR97} and the 
dimension-6 operators corresponding to them were identified. In 
particular, the dominant role of four-fermion operators was emphasized
by Vainshtein and Zakharov.\,\cite{VZ94}

Parisi's method can be used to determine the UV renormalon singularity
in QCD up to an overall constant\,\cite{BBK97} by means of 
renormalization group equations. These equations are formulated most 
easily for ambiguities or, equivalently, imaginary parts of Borel-type 
integrals. In QCD UV renormalons lie on the negative 
Borel axis and do not lead to ambiguities. It is technically convenient 
to consider instead the integral
\begin{equation}
\label{irr}
I[R](\alpha_s) = \int\limits_{0+i\epsilon}^{-t_c+i\epsilon} d t\,
e^{-t/\alpha_s}\,B[R](t) \qquad 
0< \frac{1}{\beta_0} < t_c < \frac{2}{\beta_0},
\end{equation}
given a series expansion $R$ in $\alpha_s=\alpha_s(\mu)$ 
and its Borel transform $B[R](t)$. The integral is 
complex and its imaginary part is unambiguously related to the 
first UV renormalon singularity at $t=-1/\beta_0$ or the corresponding
large-order behavior. 

The UV factorization properties of Green functions imply that 
the imaginary part of $I[R]$ owing to the leading UV renormalon 
can be represented as 
\begin{equation}
\label{parisi}
\mbox{Im}\,I[R](\alpha_s,p_k) = \frac{1}{\mu^2}\sum_i
C_i(\alpha_s)\,R_{{\cal O}_i}(\alpha_s,p_k).
\end{equation} 
In this equation ${\cal O}_i$ denote dimension-6 operators and 
$R_{{\cal O}_i}$ the Green function from which $R$ is derived with a 
single zero-momentum insertion of ${\cal O}_i$. $C_i(\alpha_s)$ 
are the coefficient functions, which are independent of any external 
momentum $p_k$ of $R$ and in fact independent of the quantity $R$. 
They play the same role as the universal renormalization constants 
in ordinary renormalization. The coefficient function being universal, 
the dependence of the UV renormalon 
divergence on the observable $R$ is contained in the factors 
$R_{{\cal O}_i}$. These factors can be computed order by order 
in $\alpha_s$ by conventional methods. We will not prove
Eq.~(\ref{parisi}) here, but rely on the fact that familiarity with 
ordinary renormalization will make it appear plausible. 
 
The dimension-6 operators 
may be thought of as an additional term,  
\begin{equation}
\Delta {\cal L} = -\frac{i}{\mu^2}\sum_i
C_i(\alpha_s)\,{\cal O}_i, 
\end{equation}
in the QCD Lagrangian with coefficients such that for {\em any} $R$ the 
imaginary part of $I[R]$ is compensated by the additional 
contribution to $R$ from $\Delta {\cal L}$. From the requirement 
that $\Delta {\cal L}$ be independent of the renormalization scale $\mu$ 
or from a comparison of the renormalization group equations satisfied by  
$I[R]$ and $R_{{\cal O}_i}$ it can be derived that 
\begin{equation}
\label{rge}
\left[\left(\beta(\alpha_s)\frac{d}{d\alpha_s} - 1\right)\delta_{i j} - 
\frac{1}{2}\gamma_{ij}(\alpha_s)\right] C_j(\alpha_s) = 0,
\end{equation}
where $\gamma(\alpha_s)$ is the anomalous dimension matrix of the 
dimension-6 operators 
defined such that the renormalized operators satisfy
\begin{equation}
\label{defandim}
\left(\delta_{ij}\,\mu\frac{d}{d\mu}+\gamma_{ij}\right)
{\cal O}_j = 0.
\end{equation}
The unusual ``$-1$'' in Eq.~(\ref{rge}) originates from the factor 
$1/\mu^2$ in Eq.~(\ref{parisi}). 
The solution to the differential equation (\ref{rge}) can be written as
\begin{equation}
\label{solcoeff}
C_i(\alpha_s) =
e^{1/(\beta_0\alpha_s)}\alpha_s^{\beta_1/\beta_0^2}\,F(\alpha_s)\,
E_i(\alpha_s),
\end{equation}
where 
\begin{equation}
\label{fi}
F(\alpha_s) = \exp\left(\,\int_0^{\alpha_s} dx\left[
-\frac{1}{\beta_0 x^2}+\frac{\beta_1}{\beta_0^2 x}-\frac{1}{\beta(x)}
\right]\right)
\end{equation}
has a regular series expansion in $\alpha_s$ and incorporates the 
effect of terms of higher order than $\beta_1$ in the $\beta$-function 
and 
\begin{equation}
\label{ei}
E_i(\alpha_s) = \exp\left(\,\int_{\alpha_0}^{\alpha_s} dx\,
\frac{\gamma_{ij}^T(x)}{2 \beta(x)}\right) \hat{C}_j
\end{equation}
takes into account the (transpose of the) 
anomalous dimension matrix. Thus, the 
coefficient functions are determined 
up to $\alpha_s$-independent 
integration constants $\hat{C}_i$. (The lower limit 
$\alpha_0$ in Eq.~(\ref{ei}) is arbitrary. A change of $\alpha_0$ can be 
compensated by adjusting the integration constants.) Because 
the $\alpha_s$-dependence in Eq.~(\ref{irr}) translates into $n$-dependence 
of large-order behavior, we deduce that this 
$n$-dependence is completely determined. Only overall normalization factors 
related to the integration constants do not follow from the renormalization 
group equation. However, these integration constants are 
process-independent numbers; they depend only on the Lagrangian that 
specifies the theory. It is in this precise sense that ultraviolet 
renormalon divergence is universal. This should be contrasted to the 
case of IR renormalons, which depend not only on the Lagrangian, but 
also on other IR parameters relevant to a specific class of
processes. 

We now specify a basis of dimension-6 operators. In general, we
are also interested in processes induced by external currents. Here  
we consider only vector and axial-vector currents and we let them be 
flavor singlets. Thus, 
in expressions like $(\bar{\psi} M\psi)$, a sum over flavor, color and 
spinor indices is implied, and $M$ is a matrix in color and spinor space, 
but unity in flavor space. 
To account for the external currents, two (abelian) 
background fields $v_\mu$ and 
$a_\mu$, which couple to the vector and axial-vector current, are introduced. 
Their fields 
strengths $F_{\mu\nu}=\partial_\mu v_\nu-\partial_\nu v_\mu$ and 
$H_{\mu\nu}=\partial_\mu a_\nu-\partial_\nu a_\mu$ satisfy 
$\partial_\mu F^{\mu\nu}=j_V^\nu$ and 
$\partial_\mu H^{\mu\nu}=j_A^\nu$. A basis of dimension-6 
operators is then given by 
\begin{eqnarray}
{\cal O}_1 &=& (\bar{\psi}\gamma_\mu\psi) (\bar{\psi}\gamma^\mu\psi)
\qquad\qquad\,\,\,
{\cal O}_2 \,=\, (\bar{\psi}\gamma_\mu\gamma_5\psi) (\bar{\psi}
\gamma^\mu\gamma_5\psi)
\nonumber\\[0.2cm]
{\cal O}_3 &=& (\bar{\psi}\gamma_\mu T^A\psi) (\bar{\psi}\gamma^\mu 
T^A \psi)
\qquad
\!\!{\cal O}_4 \,=\, (\bar{\psi}\gamma_\mu\gamma_5 T^A\psi) (\bar{\psi}
\gamma^\mu\gamma_5 T^A\psi)
\nonumber
\end{eqnarray}
\vspace*{-0.6cm}
\begin{eqnarray}
\label{basis}
{\cal O}_5 &=& \frac{1}{g_s}\,f_{ABC} \,G_{\mu\nu}^A G_\rho^{\nu\,B} 
G^{\rho\mu\,C} 
\end{eqnarray}
\vspace*{-0.6cm}
\begin{eqnarray}
{\cal O}_6 &=& \frac{1}{g_s^2}\,(\bar{\psi}\gamma_\mu\psi)\,
\partial_\nu F^{\nu\mu}
\qquad
{\cal O}_7 \,=\, \frac{1}{g_s^2}\,(\bar{\psi}\gamma_\mu\gamma_5\psi)\,
\partial_\nu H^{\nu\mu}
\nonumber\\
{\cal O}_8 &=& \frac{1}{g_s^4}\,\partial_\nu F^{\nu\mu}\,
\partial^\rho F_{\rho\mu}
\qquad\quad\!\!
{\cal O}_9 \,=\, \frac{1}{g_s^4}\,\partial_\nu H^{\nu\mu}\,
\partial^\rho H_{\rho\mu},
\nonumber
\end{eqnarray}
where the overall factors $1/g_s^k$ have been inserted for convenience. We  
neglected gauge-variant operators and operators that vanish by 
the equations of motion. We also assume that all $N_f$ quarks are 
massless. Chirality then allows us to omit four-fermion operators of 
scalar, pseudo-scalar or tensor 
type. Diagrammatically, they cannot be generated in 
massless QCD, because 
the number of Dirac matrices on any fermion line that connects 
to an external fermion in a four-point function is always odd. 

To determine the leading asymptotic behavior we only need the 
leading term in $\alpha_s$ of Eq.~(\ref{parisi}). To this order 
$F(\alpha_s)=1$ and only the leading order anomalous dimension matrix 
is needed to compute the factors $E_i(\alpha_s)$. We will not present 
this calculation\,\cite{BBK97} in detail here, but discuss 
briefly the structure of the result for the correlation function 
of two vector currents, $\Pi(Q)$. The leading contribution 
is caused by mixing of the four-quark operators into the operator 
${\cal O}_8$. Taking into account that 
$\Pi(Q)_{{\cal O}_8}(\alpha_s,Q)\propto \alpha_s^{-2}$ at leading order, 
we obtain 
\begin{equation}
\label{impiuv}
\mbox{Im}\,I[\Pi](\alpha_s,Q) \propto \frac{Q^2}{\mu^2}\,
e^{1/(\beta_0\alpha_s)}\alpha_s^{-2-\lambda_1+\beta_1/\beta_0^2} 
(1+O(\alpha_s)),
\end{equation} 
where $\lambda_1$ is related to the largest eigenvalue of the 
anomalous dimension matrix of the four-quark operators. Converting 
the previous equation into large-order behavior gives, taking
$\mu=Q$,   
\begin{equation}
\label{adleras}
r_n \stackrel{n\to \infty}{=} K\,(-\beta_0)^n\,n!\,
n^{2-\beta_1/\beta_0^2+\lambda_1} = K\,(-\beta_0)^n\,n!\,
n^{\{1.59,1.75,1.97\}},
\end{equation}
for $N_f=\{3,4,5\}$. The constant $K$ is related to the integration 
constants $\hat{C}_j$ and it is not known how to compute it. The 
result generalizes to a non-flavor-symmetric vector current with 
a different constant $K$. 

It is interesting to note that subleading corrections 
to the asymptotic behavior can be computed without introducing 
further ``non-perturbative'' parameters in addition to the constants 
$\hat{C}_i$ already present at leading order. As a rule, to obtain 
the coefficient of the $1/n^k$ correction, one needs the 
$\beta$-function coefficients $\beta_0,\ldots,\beta_{k+1}$, the 
$(k+1)$-loop anomalous dimension matrix and the $k$-loop correction 
to Green functions with operator insertions.
The renormalization group treatment can in principle be extended to the 
next singularity in the Borel plane at $t=-2/\beta_0$, but this is far more 
complicated, since 
single insertions of dimension-8 operators and double insertions 
of dimension-6 operators would have to be taken into account. 

Now that we know how to compute UV renormalons we return to the
question whether there exist power corrections associated with them. 
Equation~(\ref{impiuv}) shows that the power-suppressed imaginary part 
is, after all, polynomial in $Q$. Keeping the renormalization scale 
$\mu$ and $Q$ different, we then find that the minimal term of the 
UV renormalon series scales as\,\cite{BZ92} 
\begin{equation}
\frac{Q^2}{\mu^2}\,\frac{\Lambda^2}{\mu^2}\times[\mbox{powers of }
\alpha_s]
\end{equation}
rather than $(\Lambda/Q)^2$. Since $\mu$ is an arbitrary parameter 
we are free to make it large. The effects of this is to delay the 
onset of UV renormalon divergence, as the minimal term of the series 
occurs at order $n\sim 1/(\beta_0\alpha_s(\mu))$, and to 
simultaneously render the minimal term small. (If the number of 
known terms in the series is small, this is not a practical way 
to deal with UV renormalons. In this case one can make use of 
conformal mappings in the Borel plane to render the UV renormalon 
singularities less important.\,\cite{M92,ANR95,SS96}) 
Essentially this phenomenon occurs because the natural scale of 
the UV renormalon divergence is much larger than $Q$. By adjusting 
$\mu$ in this way, we decrease the perturbative coefficients. Since 
the strong coupling also decreases, the UV renormalon divergence 
becomes irrelevant. This is to be contrasted with the case of 
IR renormalons in which case $\mu$ would have to be adjusted 
to low values. As the perturbative coefficients decrease, the coupling
increases and both effects exactly compensate each other as far as 
the minimal term of the series is concerned.
Note that these purely perturbative 
considerations do not imply that there cannot 
be any non-perturbative power correction associated with 
short distances. 
But they tell us that whatever these corrections are they bear no 
relation with UV renormalons. 

\section{Conclusion}
\label{sec:conclusion}

During the past few years infrared renormalons have emerged as a
useful tool for the analysis of IR contributions to Feynman diagrams 
beyond the leading power in the perturbative scale, and, 
by implication, for the analysis of non-perturbative power corrections 
to generic hard processes. 

In this review we have given a brief survey of the basic ideas and 
considered a few selected applications of the method, which we think are 
the most relevant. An important problem that we have not discussed
is the relation of renormalons to lattice calculations in general
and in particular to non-perturbative lattice renormalization procedures    
for higher-twist operators. Possible non-perturbative corrections 
coming from small distances (e.g.\ due to instanton contributions) 
have also been left out.

We attempted to emphasize that the subject of IR 
renormalons is closely related to the extension 
of QCD factorization theorems on logarithmic IR singularities  
to leading power corrections in the hard 
scale. This extended factorization is at present 
well understood only at the 1-loop 
level (e.g.\ for event shapes). An outstanding problem that remains 
even at this level is
to compute the (leading) logarithmic corrections to the power behavior.
In the few cases for which the result is known, its derivation 
is based on the operator product expansion and the  
calculation of the anomalous dimensions of the relevant 
higher-twist operators. The extension to event shapes would require
an understanding of the renormalization group equations for 
the corresponding 
shape functions (cf. Sect.~\ref{sec:applications}) that have yet to 
be written down.

An important formal problem that remains is the calculation of the overall
constants of the large-order behavior. This task cannot be addressed 
within perturbation theory itself and requires new methods. Some 
recent developments\,\cite{Lee99,BB00} may be interesting in this respect, 
but it appears unlikely that an exact analytic solution can be found.

\end{document}